\documentclass[11pt,a4paper]{article}

\usepackage[margin=1in]{geometry}
\usepackage[T1]{fontenc}
\usepackage[utf8]{inputenc}
\usepackage{lmodern}
\usepackage{microtype}

\usepackage{amsmath}
\usepackage{amssymb}
\usepackage{graphicx}
\usepackage{booktabs}
\usepackage{multirow}
\usepackage{pifont}
\usepackage{subcaption}
\usepackage{xspace}
\usepackage[dvipsnames]{xcolor}
\usepackage[affil-it]{authblk}

\usepackage[colorlinks,linkcolor=MidnightBlue,citecolor=MidnightBlue,urlcolor=MidnightBlue,breaklinks]{hyperref}
\usepackage[capitalize]{cleveref}
\crefname{section}{Sec.}{Secs.}
\Crefname{section}{Section}{Sections}
\crefname{table}{Tab.}{Tabs.}
\Crefname{table}{Table}{Tables}
\crefname{appendix}{App.}{Apps.}
\Crefname{appendix}{Appendix}{Appendices}

\makeatletter
\DeclareRobustCommand\onedot{\futurelet\@let@token\@onedot}
\def\@onedot{\ifx\@let@token.\else.\null\fi\xspace}
\def\eg{\emph{e.g}\onedot} 
 
\def\cf{\emph{cf}\onedot}

\makeatother

\newcommand{\cmark}{\ding{51}}
\newcommand{\xmark}{\ding{55}}

\title{Underwater Spectral Recovery via Multi-View Hyperspectral Imaging}

\author[1]{Jonathan Sterckx\thanks{Corresponding author: \href{mailto:jonathan.sterckx@ugent.be}{jonathan.sterckx@ugent.be}}}
\author[2]{Alice James}
\author[2]{Avishkar Seth}
\author[1]{Laurens Diels}
\author[1]{Michiel Vlaminck}
\author[1]{Jan Aelterman}
\author[2]{Endrowednes Kuantama}
\author[1]{Hiep Luong}
\affil[1]{Ghent University -- imec, Belgium}
\affil[2]{Macquarie University, Australia}
\date{}

\begin{document}

\maketitle

\begin{abstract}
Underwater imaging suffers from wavelength-dependent absorption and scattering that distort scenes and obscure spectral signatures. While RGB restoration improves visual clarity, the outputs remain broadband and tied to the sensor's color space. We introduce a physically grounded framework that lifts underwater restoration to continuous spectral radiance using multi-view hyperspectral imaging. By representing continuous spectra via low-rank basis functions and leveraging multi-distance observations, our framework jointly constrains scene radiance and water parameters. We employ a robust sparse-to-dense pipeline: water parameters are estimated from consistent multi-view matches and then applied globally to invert the light transport model across the full images. Validated on physics-based synthetic data rendered via the Mitsuba 3 spectral path tracer, our method recovers physically meaningful radiance spectra and, as a downstream application, outperforms standard image restoration baselines.

\end{abstract}

\section{Introduction}
\label{sec:intro}

Underwater imaging is critical for applications like ecological monitoring~\cite{apprill2023toward, ho2020inspection}, pipeline inspection~\cite{ho2020inspection}, benthic mapping~\cite{loureiro2024survey, lowe2025benthicnet} and archaeology~\cite{costa2019underwater, huang2026application}. Some of these tasks require more than just perceptually clear images and rely on accurate spectral signatures to identify materials and infer physical properties. However, existing underwater restoration methods focus exclusively on RGB images, and operate strictly within the original color space (RGB in, RGB out). The collapse of rich spectral information into broadband channels prevents them from recovering the continuous radiance spectrum required for physically accurate analysis.

To bridge this gap, we introduce the first framework that lifts underwater restoration from discrete channels to continuous spectral radiance, recovering a full, continuous radiance spectrum for every pixel instead of performing channel-to-channel restoration. We make this underdetermined inverse problem feasible through four key design choices:
\begin{itemize}
    \item \textbf{Hyperspectral imaging (HSI):} Utilizing $C \gg 3$ narrow bands provides the spectral resolution necessary to resolve metamerism and uniquely recover radiance spectra.
    \item \textbf{Spectral modeling:} We model continuous spectra as a linear combination of basis functions, keeping the physics-based estimation low-rank and computationally efficient.
    \item \textbf{Multi-view image acquisition:} Multi-distance observations of the same points in the scene jointly constrain both scene radiance and water parameters, resolving ambiguities irreducible from a single viewpoint.
    \item \textbf{Sparse-to-dense recovery:} Water parameters are first estimated using a robust, sparse subset of multi-view points, naturally filtering out transient distractors like marine snow. These parameters are then applied globally to invert the image formation model for every pixel, while ensuring robustness against noise amplification through regularization.
\end{itemize}

While inverting the image formation model recovers the scene radiance as if no attenuation occurred along the viewing ray, the recovered radiance remains affected by depth-dependent ambient illumination. Operating in the continuous spectral domain enables our framework to perform a physically grounded white-balancing step that removes this residual chromatic bias. Furthermore, the recovered spectra can be projected onto arbitrary RGB or HSI camera response functions, facilitating both human visualization and compatibility with downstream vision models.

Since real pixel-aligned underwater ground truth is unattainable, we validate our framework using a physically rigorous rendering pipeline built on Mitsuba~3 \cite{jakob2022mitsuba3}. We evaluate spectral accuracy and robustness across diverse environmental parameters, ablate our core design choices, and benchmark our approach on a downstream image restoration task. Our results demonstrate that our method captures meaningful spectral signatures and yields clear improvements over existing image restoration methods.

The rest of this paper is organized as follows. \cref{sec:preliminaries} reviews underwater light transport, and \cref{sec:related_work} discusses related work. Our framework is detailed in \cref{sec:methods} and experimentally validated in \cref{sec:experiments}. Finally, \cref{sec:discussion} and \cref{sec:conclusion} provide a discussion and concluding remarks. Additional theoretical, methodological, and experimental details are provided in the appendices.

\section{Preliminaries}
\label{sec:preliminaries}

\begin{figure}[tb]
    \centering
    \includegraphics[width=0.99\linewidth]{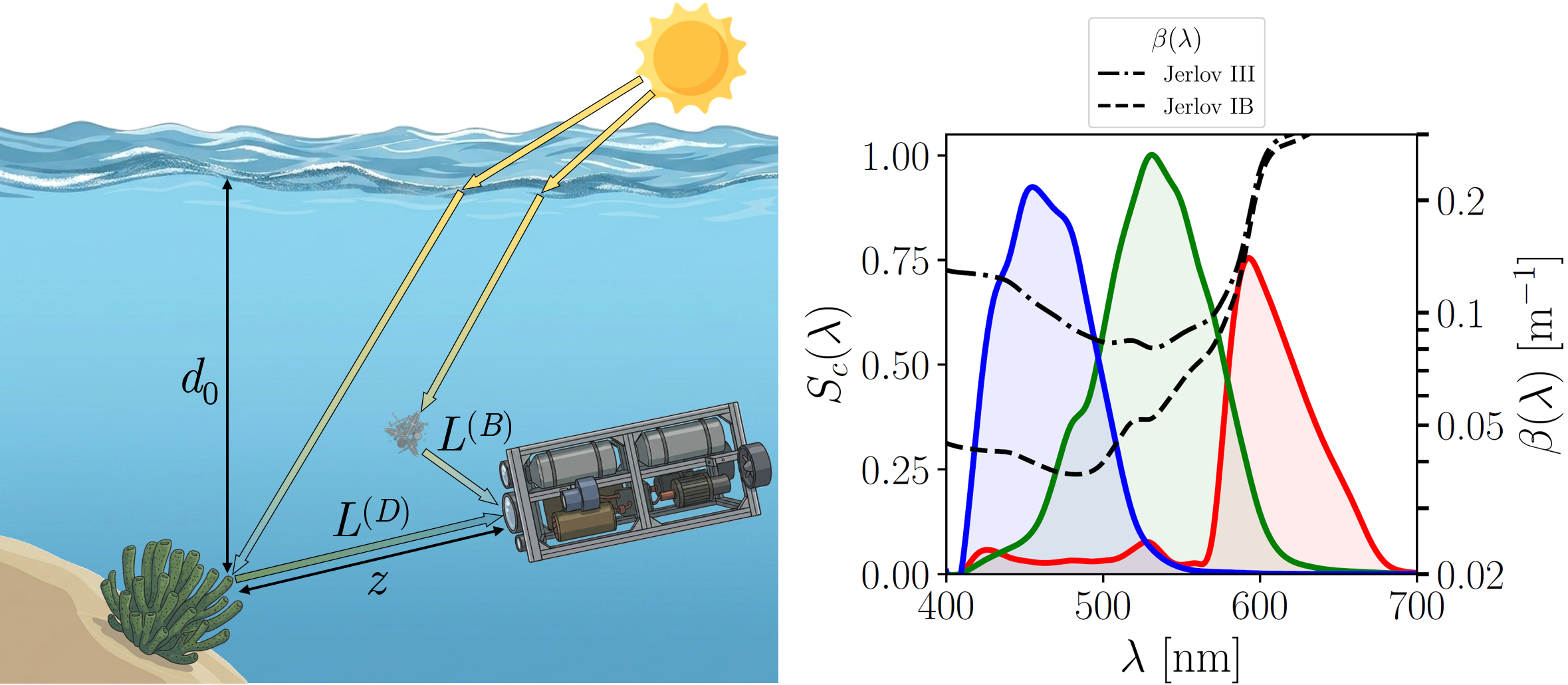}
    \caption{\textbf{Mechanisms of underwater image formation.} \textbf{Left:} Observed radiance is the sum of attenuated direct transmission $L^{(D)}$ and backscatter $L^{(B)}$. \textbf{Right:} Water optical properties, such as the beam attenuation coefficient $\beta(\lambda)$, vary strongly with wavelength (illustrated for two oceanographic water types), whereas RGB camera channels (shown for a Nikon D700) integrate over broad spectral bands, motivating hyperspectral imaging.}
    \label{fig:underwater_imaging}
\end{figure}

Underwater imaging is governed by wavelength-dependent absorption and scattering~\cite{mobley1994light}, which diverge significantly from atmospheric light transport. Building upon the formulation and notation by Akkaynak and Treibitz~\cite{akkaynak2018revised}, we review the essential components of underwater image formation to establish the physical foundation for our framework.

Consider a Lambertian surface element at depth $d_0$, observed by a sensor along a viewing ray of length $z$ (see \cref{fig:underwater_imaging}). At wavelength $\lambda$, the continuous spectral radiance $L_0(\lambda)$ emitted by the surface is a product of its intrinsic reflectance $\rho(\lambda)$ and the local ambient downwelling irradiance spectrum $E(d_0,\lambda)$:
\begin{equation} \label{eq:source_radiance}
    L_0(\lambda) = \rho(\lambda)\, E(d_0,\lambda) = \rho(\lambda)\, E(0,\lambda) e^{-K_d(\lambda) \cdot d_0},
\end{equation}
where $E(0,\lambda)$ is the atmospheric spectrum and $K_d(\lambda)$ is the water's diffuse attenuation coefficient. As $K_d(\lambda)$ is typically smallest in the blue-green portion of the spectrum, these wavelengths penetrate deepest, giving natural water its characteristic blue-green appearance.

By the Radiative Transfer Equation (RTE), the total radiance $L(\lambda, z)$ reaching the sensor is a linear combination of the attenuated \textit{direct component} $L^{(D)}$ and the particle-induced \textit{backscatter component} $L^{(B)}$ (\cref{fig:underwater_imaging}, left). Adopting the widely used horizontal viewing approximation for local depth variations~\cite{akkaynak2018revised}, the total radiance is expressed as:
\begin{equation} \label{eq:RTE_full}
    L(\lambda, z) = \underbrace{L_0(\lambda) e^{-\beta(\lambda) z}}_{L^{(D)}} + \underbrace{B^{\infty}(\lambda) \bigl(1 - e^{-\beta(\lambda)z}\bigr)}_{L^{(B)}},
\end{equation}
where $\beta(\lambda)$ is the beam attenuation coefficient and $B^{\infty}(\lambda)$ denotes the asymptotic veiling light.

An image sensor, together with its spectral filter array, discretizes the continuous incoming radiance $L(\lambda,z)$ into $C$ color channels. The measured intensity $I_c$ of channel $c$ is given by:
\begin{equation} \label{eq:pixel_value_full}
    I_c = \underbrace{\int_\Omega S_c(\lambda) L_0(\lambda)\,e^{-\beta(\lambda)z}\,\mathrm{d}\lambda}_{I_c^{(D)}} + \underbrace{\int_\Omega S_c(\lambda)B^{\infty}(\lambda)\bigl(1 - e^{-\beta(\lambda)z}\bigr) \,\mathrm{d}\lambda}_{I_c^{(B)}},
\end{equation}
where $S_c(\lambda)$ represents the spectral sensitivity of channel $c$ (\cref{fig:underwater_imaging}, right) and $\Omega$ is the sensor's spectral range.

This integration is fundamentally lossy, allowing distinct spectra to produce identical channel responses, a phenomenon known as metamerism. For conventional RGB sensors ($C = 3$), the complex, wavelength-dependent structures of $\beta(\lambda)$ and $L_0(\lambda)$ are collapsed into just three scalar values per pixel, rendering the recovery of $L_0(\lambda)$ severely underdetermined. Conversely, hyperspectral imaging cameras ($C \gg 3$) sample this profile at a higher spectral resolution, better preserving the underlying spectral signatures and making the inversion less ambiguous without overly restrictive statistical priors.

If the water along the optical path between camera and object were evacuated ($\beta \to 0$), the sensor would register the \emph{latent intensity} $J_c$:\footnote{While $J_c$ is free from line-of-sight attenuation, it is not fully ``water-free'' because $L_0(\lambda)$ still retains the depth-dependent chromatic bias of the downwelling light $E(d_0,\lambda)$ (see \cref{eq:source_radiance}). To compensate, subsequent white balancing is required.}
\begin{equation} \label{eq:latent_image}
    J_c = \int_\Omega S_c(\lambda)\,L_0(\lambda)\,\mathrm{d}\lambda.
\end{equation}

Instead of relying on conventional RGB approximations that simplify \cref{eq:pixel_value_full} into error-prone, channel-wise effective constants (detailed in \cref{app:broadband}), our framework explicitly preserves the continuous spectral domain. We solve directly for the intrinsic physical variables $L_0(\lambda)$, $\beta(\lambda)$, and $B^\infty(\lambda)$, using HSI to resolve the narrow-band interactions.

\section{Related Work}
\label{sec:related_work}

In this section, we briefly review the core underwater image restoration paradigms to highlight the gaps our continuous spectral recovery framework bridges.

\textbf{Single-view} image enhancement approaches are broadly categorized into heuristics, deep learning, and physics-based inversion. Heuristics like CLAHE~\cite{zuiderveld1994contrast} or Dark Channel Prior~\cite{he2010single} introduce artificial color shifts and lack physical grounding. Deep learning methods~\cite{li2021underwater, wu2021two, chen2024uwformer} improve perceptual quality but prioritize visual appearance over radiometric accuracy, limiting generalization. Physics-based frameworks, most notably Sea-thru~\cite{akkaynak2019sea}, invert the image formation model using estimated distance maps and veiling light. Despite different formulations, these methods all restore images within the original sensor measurement space (RGB $\to$ RGB), rather than recovering the underlying continuous spectral radiance.

\textbf{Multi-view} frameworks exploit redundant viewpoints to observe identical 3D points at varying distances $z$, providing explicit constraints to optimize water parameters without brittle single-view heuristics. Recently, physics-based models have been integrated into differentiable volumetric pipelines like NeRF~\cite{levy2023seathru, sethuraman2023waternerf} and 3D Gaussian Splatting~\cite{yang2025seasplat, zhang2026waterclear}, to decouple latent scene appearance from medium degradation.

However, these methods suffer from two key limitations. Operationally, their heavy reliance on dense photometric consistency makes them highly sensitive to dynamic marine environments (\eg, moving sea life, marine snow, changing caustics), which introduce reconstruction artifacts and require costly scene-specific optimization~\cite{gough2025aquanerf}. Physically, they remain bound to broadband approximations, inheriting the systematic errors of conflated, sensor-dependent variables.

\textbf{Hyperspectral} sensors provide the fine spectral resolution needed to uniquely characterize materials, but physics-based underwater HSI restoration remains largely unexplored. Existing model-based spectral methods often rely on impractical calibration steps, such as imaging the identical scene both in air and underwater~\cite{guo2016model}. Consequently, most underwater HSI pipelines bypass restoration entirely, directly feeding uncorrected, degraded spectral textures into downstream mapping or classification tasks~\cite{ferrera2021hyperspectral, mills2023assessment}. Active systems like Hyperspectral LiDAR~\cite{wang2025feasibility} achieve radiometric correction but require specialized hardware and lack the resolution of passive sensors.

\textbf{A critical gap} exists at the intersection of physical light transport, multi-view geometry, and hyperspectral imaging. Existing HSI workflows rely on laboratory-bound air calibration or ignore water degradation during 3D mapping. Conversely, existing restoration methods remain confined to sensor-specific channel spaces, producing restored RGB observations rather than recovering the underlying spectral radiance. Volumetric approaches further rely on broadband approximations and monochromatic assumptions. Our framework is the first to bridge this gap by embedding a spectrally continuous image formation model directly into a multi-view pipeline. This enables joint optimization of spectral radiance $L_0(\lambda)$ and intrinsic medium parameters, free from channel-wise proxies and static scene assumptions.

\section{Methods}\label{sec:methods}

We present a physically grounded optimization framework to recover continuous spectral scene radiance from multi-view hyperspectral images.

\subsection{Inverse Problem and Spectral Representation}\label{sec:methods_formulation}

Our objective is to recover the spectral radiance $L_0(\lambda)$ of scene points from images with $C$ spectral channels under uniform ambient illumination and Lambertian surface assumptions.\footnote{We discuss the implications of these assumptions in \cref{sec:discussion}.}

The image formation process (\cref{eq:pixel_value_full}) maps scene, medium and sensor properties to a measured intensity:
\begin{equation*}
    \big\{ L_0(\lambda), z, \beta(\lambda), B^\infty(\lambda), S_c(\lambda) \big\} \mapsto I_c.
\end{equation*}
The channel sensitivity $S_c(\lambda)$ is assumed known \textit{a priori}, while the range $z$ is obtained via ranging sensors, stereo vision, photogrammetry, or monocular depth estimation.

The medium properties $\beta(\lambda)$ and $B^\infty(\lambda)$ are spatially uniform while the intrinsic radiance $L_0(\lambda)$ is unique to each scene point. Hence, for any point, variations in the measured intensity $I_c$ across multiple views arises solely from changes in $z$. Observing multiple points, each across various distances, provides the constraints needed to jointly estimate medium attenuation and isolate point-wise radiance.

To resolve this high-dimensional, ill-conditioned inverse problem efficiently, we project the continuous spectra onto low-dimensional, compact bases, motivated by the inherent compressibility of natural spectra~\cite{cohen1964dependency, vrhel1994measurement}.
\begin{equation}\label{eq:radiance_basis}
L_0(\lambda) \approx \sum_{k=1}^{K_L} w_k^{(L)} \phi_k^{(L)}(\lambda), \,\,\,\, \beta(\lambda) \approx \sum_{k=1}^{K_\beta} w_k^{(\beta)} \phi_k^{(\beta)}(\lambda), \,\,\,\, B^\infty(\lambda) \approx \sum_{k=1}^{K_B} w_k^{(B)} \phi_k^{(B)}(\lambda),
\end{equation}
where $\phi(\lambda)$, $w$ and $K$ denote the respective basis functions, optimization coefficients, and basis dimensionality.
This dimensionality reduction reduces the parameter space, stabilizes optimization, and suppresses high-frequency noise. We evaluated both data-driven and cubic B-spline bases, and empirically found B-splines to yield more stable optimization (\cf \cref{app:spectral_bases}).

\subsection{Recovery Pipeline}\label{sec:methods_core_ideas}

We split our pipeline into a sparse stage (robust water parameter estimation from stable multi-view observations) and a dense stage (full-image radiance recovery), using only informative, physically consistent observations to estimate the water parameters before restoring the complete image.

\subsubsection{Collecting and Filtering Multi-View Observations}\label{sec:methods_filtering}

\begin{figure}[tb]
    \centering
    \includegraphics[width=0.99\linewidth]{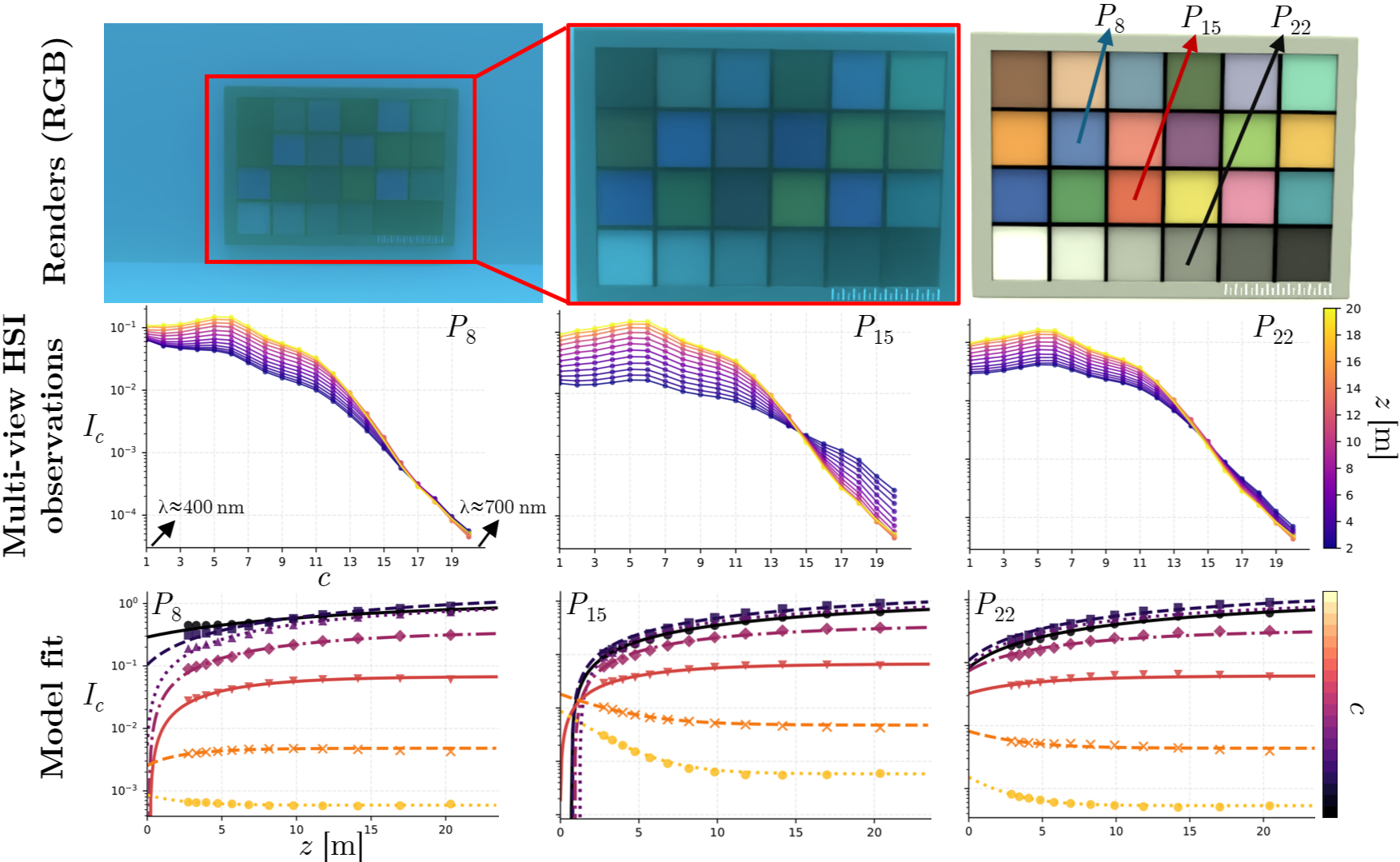}
    \caption{\textbf{Sparse recovery from multi-view observations.} \textbf{Top:} Underwater HSI image (displayed in RGB for visualization only), its clear-air reference, and three selected points ($P_8, P_{15}, P_{22}$). \textbf{Middle:} Multi-view HSI observations for each point. Each curve shows the intensity across HSI channels ($c$) for a single view $j$ at a specific line-of-sight distance $z_{ij}$; curve colors encode the distance $z$. \textbf{Bottom:} Fitted model predictions (\textit{curves}) overlaid on the discrete observations (\textit{markers}) plotted as a function of distance $z$. Here, each curve tracks a single HSI channel, with colors encoding the channel index $c$. This row visualizes the optimization alignment, showing how the multi-view data constrain both water parameters and radiances.}
    \label{fig:sparse_restoration_demo}
\end{figure}

Our framework operates on a sparse set of keypoint matches tracked across multiple views, paired with corresponding line-of-sight distances derived from Structure from Motion (SfM) or ranging sensors. For each point $P_i$ observed across a subset of images indexed by $\mathcal{V}_i \subseteq \{1, 2, \dots, N_\text{images}\}$, we extract the corresponding multi-view observations $\left\{ (\mathbf{I}_{ij}, z_{ij}) \mid j \in \mathcal{V}_i \right\}$ pairing the channel intensities $\mathbf{I}_{ij} \in \mathbb{R}^C$ with line-of-sight distances $z_{ij}$ (visualized in the middle row of \cref{fig:sparse_restoration_demo}).

To guarantee stable radiometric optimization, a point is removed if it meets any of these geometric or texture criteria (detailed in \cref{app:obs_filtering}):
(1) \textit{insufficient distance diversity}, where fewer than $\tau_\text{bins}$ distance bins of width $\delta_z$ are occupied within a valid range of $z < \tau_z$;
(2) \textit{high geometric or tracking uncertainty}, where the point's localization or distance measurement error exceeds $\tau_{\epsilon}$; or
(3) \textit{high local image contrast} ($ > \tau_c$) to mitigate pixel-mixing edge artifacts.

\subsubsection{Sparse Recovery}\label{sec:methods_sparse}

From the filtered set of multi-view observations, we jointly estimate the parameter set $\boldsymbol{\Theta} = \left\{ \mathbf{w}^{(\beta)}, \mathbf{w}^{(B)}, \{ \mathbf{w}_i^{(L)} \}_{i=1}^{N_\text{pts}} \right\}$, which comprises the water coefficients, and the radiance coefficients for all $N_\text{pts}$ points that survived the filtering step. We compute the optimal parameters $\hat{\boldsymbol{\Theta}}$ via a joint non-linear least-squares optimization:
\begin{equation}\label{eq:sparse_objective}
\hat{\boldsymbol{\Theta}}
=
\arg\min_{\boldsymbol{\Theta}}
\sum_{i=1}^{N_\text{pts}}\sum_{j\in\mathcal V_i}
\omega_{ij}
\sum_{c=1}^C
\mathcal{L}_\delta\!\left(
w_c(I_{ij,c}-\hat I_{ij,c}(\boldsymbol{\Theta}))
\right),
\end{equation}
where $\omega_{ij}$ is a dynamic outlier rejection weight, $\mathcal{L}_\delta$ is the Huber loss, and $w_c \propto \left(\sum_{i,j} |I_{ij, c}|\right)^{-1}$ represents global channel-whitening weights to prevent high-intensity channels (\eg, blue) from disproportionately dominating the loss. Inside the loss function, $I_{ij,c}$ is the observed intensity and $\hat{I}_{ij,c}(\boldsymbol{\Theta})$ is the forward-model prediction combining \cref{eq:pixel_value_full} and \cref{eq:radiance_basis}.

While the initial filtering handles static geometric errors, the dynamic weights $\omega_{ij}$ specifically suppress transient artifacts (\eg, moving caustics), image noise, or incorrect tracks. We resolve this optimization problem efficiently using a Levenberg-Marquardt framework. Details on the iterative outlier suppression scheme and the acceleration of the optimization via Schur complement marginalization are provided in \cref{app:outlier,app:schur}.

\subsubsection{Dense Recovery}\label{sec:methods_dense_restoration}

\begin{figure}[tb]
    \centering
    \includegraphics[width=0.99\linewidth]{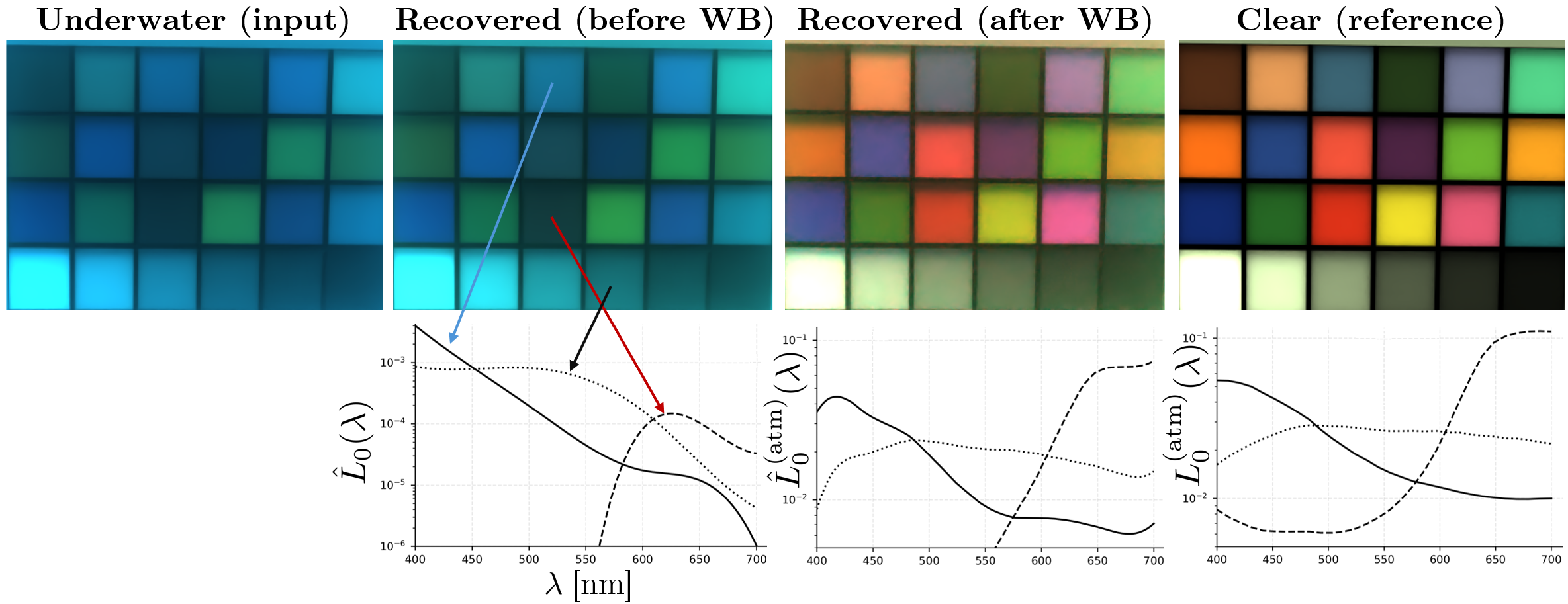}
    \caption{\textbf{Dense recovery.} \textbf{Top:} HSI renders (displayed in RGB for visualization only) of the underwater input, restored image before and after white balancing, and the corresponding clear image. \textbf{Bottom:} Recovered and corresponding clear spectra for three example points.}
    \label{fig:fig_dense_restoration_demo}
\end{figure}

With the estimated water coefficients ($\hat{\beta}(\lambda), \hat{B}^\infty(\lambda)$) fixed, recovering scene radiance across all pixels and images reduces to independent, parallelizable pixel-wise inversions of the forward model, with $L_0(\lambda)$ as the sole unknown per pixel. To ensure numerical stability in low-signal regimes (\eg, severe deep-water red attenuation) while maintaining computational efficiency, we use a multi-step decomposition. Since all pixels and images are treated independently, the following equations are written for a single, arbitrary pixel.

\paragraph{Step 1: Channel-space restoration.} The primary objective of this stage is to recover the latent intensity $\hat{J}_c$ (\cref{eq:latent_image}), which serves as the foundation for the subsequent spectral lifting. For each pixel with channel-wise intensity $I_c$ and line-of-sight distance $z$, we first isolate the estimated direct signal component $\hat{I}_c^{(D)}$ by evaluating and subtracting the background backscatter component $\hat{I}_c^{(B)}$ from the full forward model (\cref{eq:pixel_value_full}) while enforcing a non-negativity constraint:
\begin{equation} \label{I_D_hat}
\hat{I}_c^{(D)} = \max \left( I_c - \hat{I}_c^{(B)}, \, 0 \right),
\end{equation}
where $\hat{I}_c^{(B)} = \int_{\Omega} S_c(\lambda) \hat{B}^{\infty}(\lambda) \left( 1 - e^{-\hat{\beta}(\lambda) z} \right) \mathrm{d}\lambda$.

Assuming the medium transmission varies minimally within the narrow bandwidth of channel $c$, \cref{I_D_hat} simplifies to $\hat{I}_c^{(D)} \approx J_c \hat{T}_c$, where $J_c$ represents the latent intensity (\cref{eq:latent_image}) and $\hat{T}_c = \int_{\Omega} S_c(\lambda) e^{-\hat{\beta}(\lambda)z} \mathrm{d}\lambda$ the effective channel transmissivity.\footnote{This assumption introduces significant errors for broadband sensors, but holds well for narrow HSI channels. Residual modeling errors are mitigated during the spectral lifting stage.} The derivation is provided in \cref{app:transmission_derivation}.

To prevent catastrophic noise amplification where $\hat{T}_c \rightarrow 0$, we estimate $J_c$ via a regularized inversion:
\begin{equation}\label{eq:tikhonov_channel_bootstrap}
\hat J_c = \frac{\hat{T}_c}{\hat{T}_c^2 + \sigma^2} \, \hat{I}^{\mathrm{D}}_c,
\end{equation}
where $\sigma^2$ controls the Tikhonov regularization strength and is chosen according to the estimated sensor noise level.

\paragraph{Step 2: Spectral lifting.}
We lift latent channel intensities $\hat{J}_c$ into the continuous spectral space by substituting the radiance basis representation $\hat{L}_0(\lambda) = \sum_{k=1}^{K_L} \hat{w}^{(L)}_k \phi_k^{(L)}(\lambda)$ into the latent image formation model (\cref{eq:latent_image}). This discretizes the continuous forward model into a linear system $\mathbf{\hat{J}} = \mathbf{M}\hat{\mathbf{w}}^{(L)}$, where the system matrix $\mathbf{M} \in \mathbb{R}^{C \times K_L}$ maps continuous basis functions to discrete channels via $M_{c,k} = \int_{\Omega} S_c(\lambda)\phi_k^{(L)}(\lambda)\,\mathrm{d}\lambda$.

Depending on the sensor configuration, $\mathbf{M}$ can be poorly conditioned. To stabilize inversion and suppress noise amplification, we recover the radiance coefficients via ridge regression:
\begin{equation}\label{eq:dense_pixel_ridge_problem}
\hat{\mathbf{w}}^{(L)} = \arg\min_{\mathbf{w}^{(L)}} \left\| \hat{\mathbf{J}} - \mathbf{M}\mathbf{w}^{(L)} \right\|_2^2 + \xi \left\| \mathbf{w}^{(L)} \right\|_2^2,
\end{equation}
where $\xi$ is a pixel-specific penalty that varies dynamically based on local signal quality (detailed in \cref{app:adaptive_reg}). The final non-negative scene spectrum is recovered as:
\begin{equation}
\hat L_0(\lambda) = \max\left( \sum_{k=1}^{K_L} \hat{w}^{(L)}_k \phi^{(L)}_k(\lambda),\; 0 \right).
\end{equation}

\paragraph{Step 3: White balancing.}
Because the recovered radiance $\hat{L}_0(\lambda)$ retains the depth-dependent chromatic bias of the downwelling light, an optional white balancing step performs atmospheric relighting to recover $\hat{L}^{(\text{atm})}_0(\lambda)$ or isolates the intrinsic reflectance profile $\rho(\lambda)$ via
\begin{equation}
    \hat{L}^{(\text{atm})}_0(\lambda) = \frac{\hat{L}_0(\lambda) \, E(0,\lambda)}{E(d_0,\lambda)}, \qquad \rho(\lambda) = \frac{\hat{L}_0(\lambda)}{E(d_0,\lambda)}.
\end{equation}

The illumination ($E$) can be measured via an \textit{in situ} radiometer, estimated using reference calibration tiles, or approximated through heuristic gray-world priors~\cite{ma2019underwater} (see \cref{app:white_balancing}).

Optionally, the recovered spectra can be relit under arbitrary illuminants, or projected onto any channel space (\eg, RGB or custom HSI configurations) by providing custom sensitivities $\{S_c(\lambda)\}_{c=1}^C$ and applying~\cref{eq:latent_image} for human visualization or downstream neural networks trained on specific channel spaces.

\cref{fig:fig_dense_restoration_demo} illustrates an example of the dense recovery stage.

\section{Validation}
\label{sec:experiments}

We evaluate our method across spectral accuracy, environmental robustness, module ablations, and downstream image restoration benchmarks.

\subsection{Synthetic Data Generation} \label{subsec:synth_data_generation}

\begin{figure}[tb]
\centering
\begin{subfigure}{0.49\textwidth}
    \includegraphics[width=\textwidth]{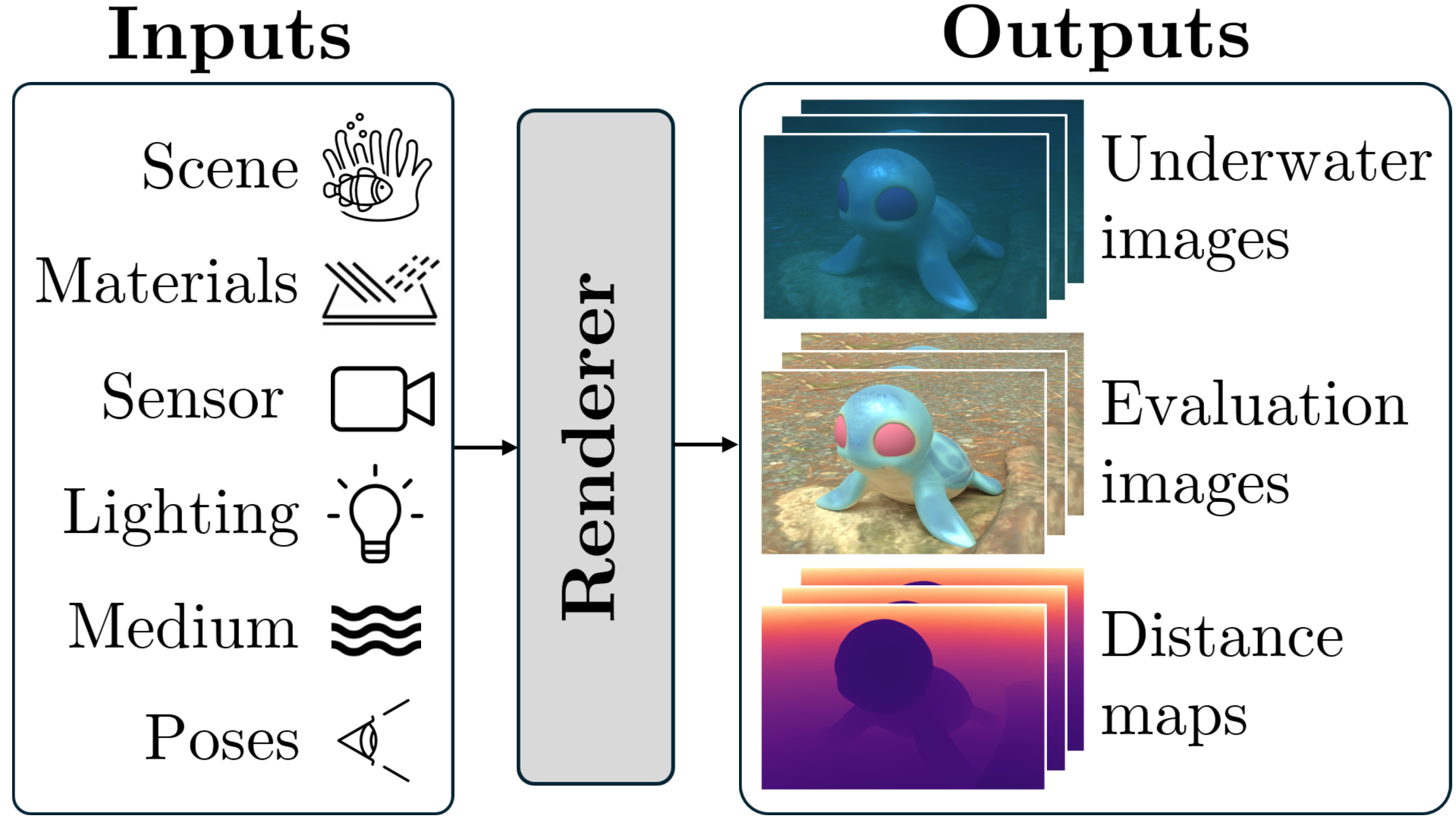}
    \caption{\textbf{Rendering inputs and outputs.} Images are displayed in RGB for visualization only.}
    \label{fig:rendering}
\end{subfigure}
\hfill
\begin{subfigure}{0.49\textwidth}
    \includegraphics[width=\textwidth]{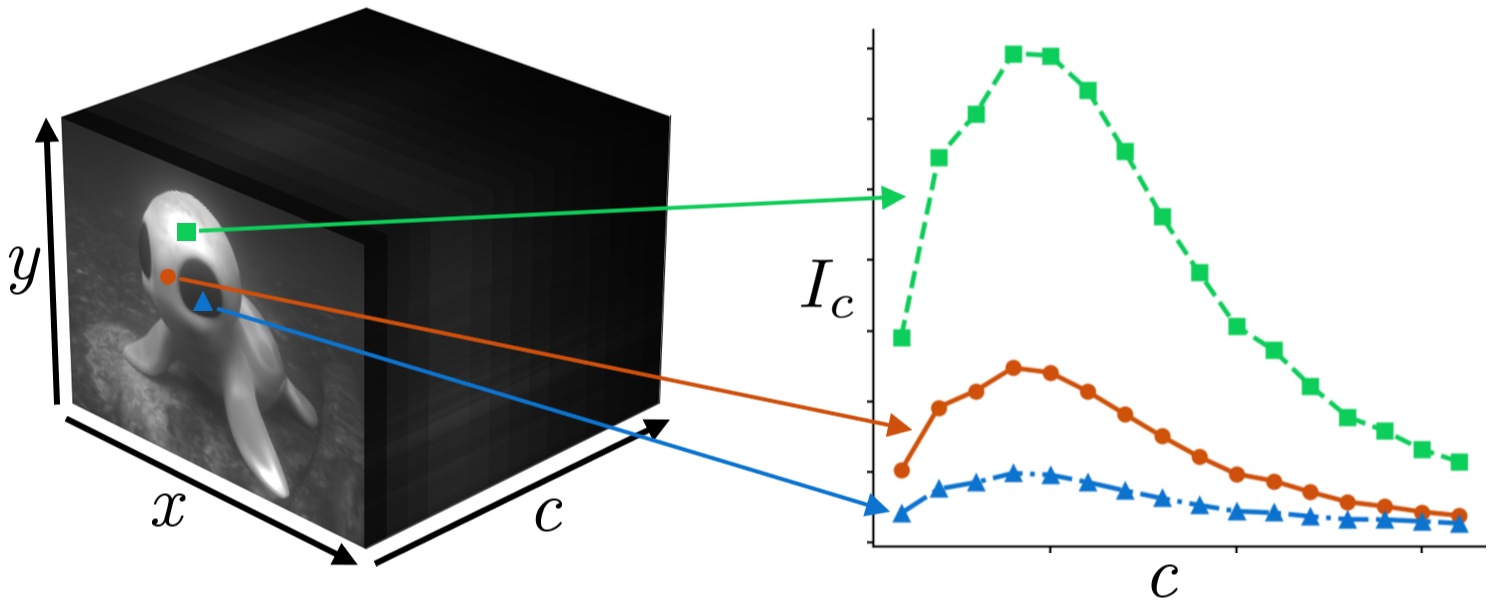}
    \caption{\textbf{Hyperspectral rendering in Mitsuba 3.} Left: rendered hyperspectral cube. Right: HSI profiles of selected pixels.}
    \label{fig:hsi_rendering}
\end{subfigure}
\caption{Synthetic data generation overview.}
\label{fig:synthetic_data_generation}
\end{figure}

Real-world underwater datasets with pixel-aligned, unattenuated ground-truth radiance spectra do not exist due to the practical acquisition challenges. To avoid testing on our own forward model or simplified closed-form models (\eg, Jaffe--McGlamery~\cite{mcglamery1980computer}), we evaluate our method using Mitsuba~3~\cite{jakob2022mitsuba3}, a physics-based renderer that numerically solves the radiative transfer and volume rendering equations via spectral path tracing. This allows us to simulate wavelength-dependent absorption, multiple scattering, and wave caustics, providing a physically grounded proxy for quantitative evaluation.

For each scene, the pipeline renders an underwater HSI image ($\in \mathbb{R}^{H \times W \times C}$), a distance map ($\in \mathbb{R}^{H \times W}$), and an atmospheric reference image ($\in \mathbb{R}^{H \times W \times V}$) where the water medium is replaced by clear air. These reference images, captured via a virtual sensor with contiguous $5$~nm bands, serve as the discretized ground-truth radiance spectra for evaluation (\cref{fig:synthetic_data_generation}).
Details on the rendering pipeline and sampling strategies are provided in \cref{app:synth_data_gen,app:mv_collection}.

\subsection{Evaluation Metrics} \label{subsec:metrics}
Let $\mathbf{L}^{(\text{atm})}, \hat{\mathbf{L}}^{(\text{atm})} \in \mathbb{R}^{H \times W \times V}$ denote the ground-truth and recovered atmospheric spectral cubes. Because the ground truth is simulated in an atmospheric environment, the recovered radiance must similarly be converted to atmospheric illumination to enable a valid comparison.\footnote{To isolate the evaluation to the core recovery pipeline rather than the quality of the illumination prior, white balancing is performed by fitting an illumination profile $E(\lambda)$ that optimally maps the recovered spectral cubes to the ground truth.} The radiance spectrum $L_0(\lambda)$ at each spatial pixel $(h, w)$ is discretized into $V=60$ contiguous $5$ nm bins spanning the $400\text{--}700$ nm range.

We evaluate performance using the normalized \textit{Mean Absolute Error} (MAE) as the average pixel-wise absolute discrepancy across all spectral channels. We also report the \textit{Spectral Angle Mapper} (SAM) to measure the average angular deviation (in degrees) between the spectral vectors:
\begin{equation}
\text{SAM} = \frac{1}{H W} \sum_{h=1}^{H} \sum_{w=1}^{W} \arccos \left( \frac{\langle \mathbf{L}^{(\text{atm})}(h,w, \cdot), \hat{\mathbf{L}}^{(\text{atm})}(h,w, \cdot) \rangle}{\|\mathbf{L}^{(\text{atm})}(h,w, \cdot)\|_2 \,\, \|\hat{\mathbf{L}}^{(\text{atm})}(h,w, \cdot)\|_2} \right).
\end{equation}
Due to severe attenuation in the $600\text{--}700$ nm range, this band degrades rapidly. To prevent it from skewing evaluation, we also report wavelength-bounded variants ($\text{MAE}_{\lambda}$ and $\text{SAM}_{\lambda}$) restricted strictly to the $400\text{--}600$ nm range.

We provide a complete list of hyperparameter configurations for all experiments in \cref{app:hyperparams}.

\subsection{Experiments}

\subsubsection{Accuracy and Sensitivity Analysis}\label{sec:sensitivity}
To evaluate the robustness of our pipeline under physical variations, we hold the method's hyperparameters fixed and systematically stress-test individual environmental parameters. This protocol utilizes a controlled synthetic \textit{Chart} scene (visualized in \cref{fig:sparse_restoration_demo}), consisting of a 24-color calibration chart against a uniform gray backdrop. The camera approaches the chart from $20\,\text{m}$ to $2\,\text{m}$ in 12 steps.

By varying individual parameters against this baseline, we analyze how different color channels are recovered under shifting medium properties, geometric constraints, and sensor constraints. Specifically, we evaluate variations across depth, water type, image noise (varying shot and Gaussian noise), distance error (varying tracking precision and outlier degradation), distance diversity (varying spatial density vs. range truncation), and channel count ($C$). The empirical findings are summarized in \cref{tab:sensitivity_study}. The configurations are detailed in \cref{app:sensitivity}.

\begin{table}[tb]
\centering
\caption{\textbf{Sensitivity study results.} We evaluate performance variations across distinct levels of physical and environmental degradations. Reference configuration (\textit{top row}): 5.0\,m depth, Jerlov IA water type, and $C=20$ channels.}
\label{tab:sensitivity_study}
\small
\setlength{\tabcolsep}{5pt}
\begin{tabular}{llcccc}
    \toprule
    \textbf{Dimension} & \textbf{Variant} & \textbf{MAE} & \textbf{MAE}$_\lambda$ & \textbf{SAM} & \textbf{SAM}$_\lambda$ \\
    \midrule
    & \textit{Reference} & 0.0772 & 0.0714 & 16.17 & 12.31 \\
    \midrule
    \multirow{2}{*}{Depth}
    & 2.5\,m  & 0.0738 & 0.0686 & 16.31 & 12.83 \\
    & 10.0\,m & 0.0837 & 0.0714 & 20.72 & 12.31 \\
    \midrule
    \multirow{2}{*}{Water Type}
    & Jerlov II & 0.1205 & 0.1110 & 21.98 & 14.94 \\
    & Clear     & 0.0101 & 0.0064 & 8.78  & 7.40 \\
    \midrule
    \multirow{3}{*}{Image Noise}
    & Conf. 1 & 0.0863 & 0.0716 & 25.25 & 12.59 \\
    & Conf. 2 & 0.0872 & 0.0723 & 22.23 & 12.58 \\
    & Conf. 3 & 0.1371 & 0.1335 & 21.17 & 15.33 \\
    \midrule
    \multirow{3}{*}{Distance Errors}
    & Conf. 1 & 0.0871 & 0.0724 & 22.67 & 13.05 \\
    & Conf. 2 & 0.0940 & 0.0748 & 25.22 & 14.34 \\
    & Conf. 3 & 0.1048 & 0.1023 & 21.76 & 16.06 \\
    \midrule
    \multirow{2}{*}{Distance Diversity}
    & Conf. 1 & 0.0828 & 0.0733 & 20.05 & 12.66 \\
    & Conf. 2 & 0.0844 & 0.0699 & 21.71 & 12.35 \\
    \midrule
    \multirow{2}{*}{Channels ($C$)}
    & 3  & 0.1111 & 0.1103 & 28.87 & 23.45 \\
    & 10 & 0.1097 & 0.1080 & 19.17 & 13.78 \\
    \bottomrule
\end{tabular}
\end{table}

\subsubsection{Ablation Study}
To validate our design choices, we conduct an ablation study systematically isolating the individual contributions of filtering, regularization, and basis decomposition. To complement the controlled chart scene experiments, we evaluate these variants across two more complex scenes (visualized in \cref{fig:qualitative_restoration_benchmarks}). The first environment models favorable conditions with clear water and no distractors to establish baseline performance in optimal scenarios. The second introduces highly attenuating water and dynamic distractors to stress-test the framework under challenging constraints. Quantitative results across both scenes are reported in \cref{tab:ablation_study}, and scene details are provided in \cref{app:scenes}.

\begin{table}[tb]
\centering
\caption{\textbf{Ablation study results.} \textit{Filt.} denotes the multi-view observation filter, \textit{Reg.} indicates the Tikhonov regularization term $\sigma^2$ in \cref{eq:tikhonov_channel_bootstrap}, and \textit{Decomp.} represents the explicit basis decomposition versus direct dense sampling.}
\label{tab:ablation_study}
\small
\setlength{\tabcolsep}{5pt}
\begin{tabular}{ccccccccc}
    \toprule
    \multicolumn{3}{c}{\textbf{Active Modules}} & & & & & & \\
    \cmidrule(r){1-3}
    \textbf{Filt.} & \textbf{Reg.} & \textbf{Decomp.} & & \textbf{Scene} & \textbf{MAE} & \textbf{MAE}$_\lambda$ & \textbf{SAM} & \textbf{SAM}$_\lambda$ \\
    \midrule
    \multirow{2}{*}{\cmark} & \multirow{2}{*}{\cmark} & \multirow{2}{*}{\cmark} & & 1 & 0.0448 & 0.0373 & 11.10 & 6.24 \\
    & & & & 2 & 0.0935 & 0.0751 & 16.44 & 7.72 \\
    \midrule
    \multirow{2}{*}{\xmark} & \multirow{2}{*}{\cmark} & \multirow{2}{*}{\cmark} & & 1 & 0.0869 & 0.0767 & 22.61 & 15.43 \\
    & & & & 2 & 0.1946 & 0.1318 & 21.80 & 9.03 \\
    \midrule
    \multirow{2}{*}{\cmark} & \multirow{2}{*}{\xmark} & \multirow{2}{*}{\cmark} & & 1 & 0.0494 & 0.0367 & 15.60 & 6.77 \\
    & & & & 2 & 0.2160 & 0.1197 & 35.49 & 17.94 \\
    \midrule
    \multirow{2}{*}{\cmark} & \multirow{2}{*}{\cmark} & \multirow{2}{*}{\xmark} & & 1 & 0.1019 & 0.0973 & 29.34 & 31.74 \\
    & & & & 2 & 0.2403 & 0.2064 & 39.36 & 39.62 \\
    \cmidrule{1-9}
    \multirow{2}{*}{\cmark} & \multirow{2}{*}{\cmark} & \multirow{2}{*}{\xmark\textsuperscript{\dag}} & & 1 & 0.0598 & 0.0517 & 12.51 & 8.53 \\
    & & & & 2 & 0.2168 & 0.1494 & 29.90 & 15.95 \\
    \bottomrule
    \addlinespace
    \multicolumn{9}{l}{\textsuperscript{\dag}\footnotesize{Utilizes Tikhonov regularization.}}
\end{tabular}
\end{table}

\subsubsection{Additional Experiments}

\begin{figure}[tb]
    \centering
    \includegraphics[width=0.99\linewidth]{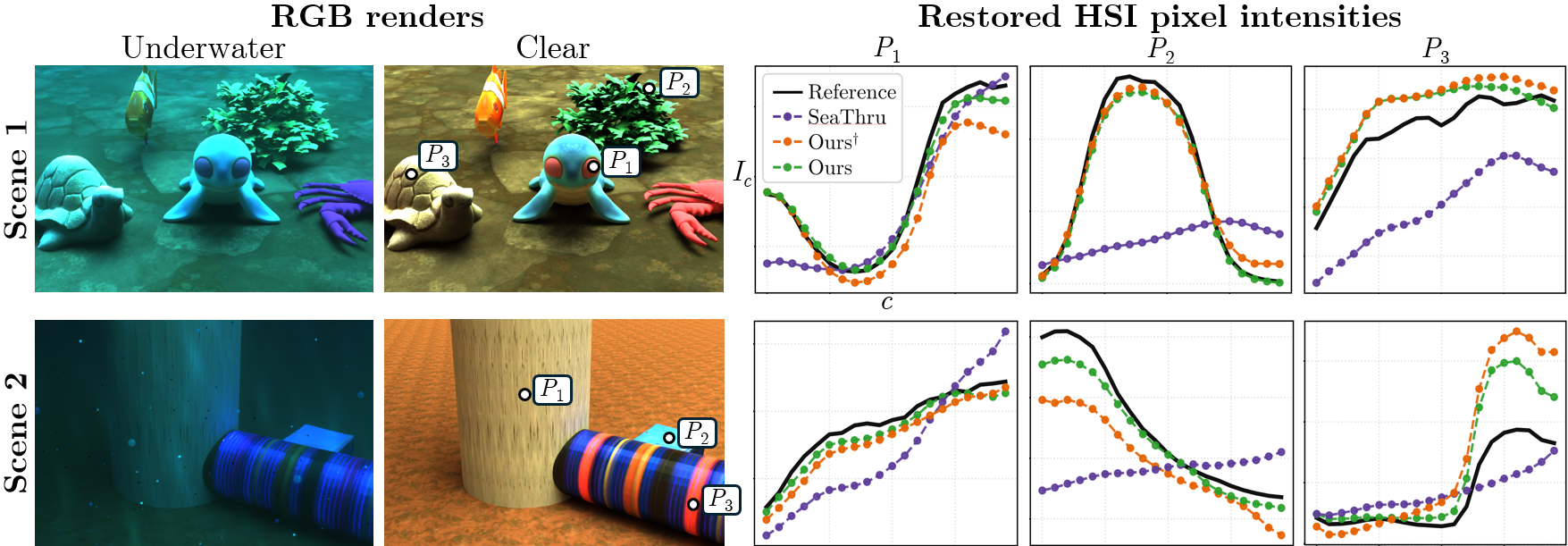}
    \caption{\textbf{Pixel-level HSI restoration showcase.} \textbf{Left:} Underwater and clear-air renders (RGB for visualization) across two scenes, highlighting selected points. \textbf{Right:} Restored HSI pixel channel intensities at the selected points. The curves compare the clear reference against profiles restored from HSI renders using different methods.}
    \label{fig:qualitative_restoration_benchmarks}
\end{figure}

Currently, no existing methods perform continuous spectral recovery. Nevertheless, we benchmark our model's performance on a downstream image restoration task to facilitate comparison with alternative paradigms. Restored images are obtained by projecting the recovered radiances back into the sensor's channel space by integrating them against the sensor sensitivity functions.
We compare our method against a channel-wise baseline variant ($\text{Ours}^\dagger$) and a hyperspectral extension of Sea-Thru~\cite{akkaynak2019sea}. The baselines and experimental configurations are detailed in \cref{app:img_restoration}. Quantitative results across the benchmark scenes are detailed in \cref{tab:scene_comparison}. For qualitative evaluation, \cref{fig:qualitative_restoration_benchmarks} showcases a preview of pixel-level spectral restorations, with full-size restored images provided in \cref{fig:restoration_scene1,fig:restoration_scene2}.

\begin{table}[tb]
\centering
\caption{Image restoration MAE of Sea-Thru, Ours$^\dagger$ and Ours across different scenes and modalities.}
\label{tab:scene_comparison}
\small
\setlength{\tabcolsep}{5pt}
\begin{tabular}{llccc}
    \toprule
    \textbf{Scene} & \textbf{Modality} & \textbf{Sea-Thru} & \textbf{Ours}$^\dagger$ & \textbf{Ours} \\
    \midrule
    \multirow{2}{*}{1} & HSI & 0.1103 & 0.0262 & 0.0186 \\
                       & RGB & 0.1038 & 0.0556 & 0.0509 \\
    \midrule
    \multirow{2}{*}{2} & HSI & 0.1012 & 0.1273 & 0.0799 \\
                       & RGB & 0.1016 & 0.0876 & 0.0982 \\
    \bottomrule
\end{tabular}
\end{table}

To address practical deployment scenarios, \cref{app:practical} briefly discusses and presents preliminary results for our pipeline when loosening the assumption of oracle data availability. Specifically, we show a representative example utilizing vision-based distance estimation rather than ground-truth geometry, alongside an overview of alternative, reference-free white balancing strategies.

\section{Discussion} \label{sec:discussion}

\subsection{Experimental Results}

Across the experiments, the reconstructed profiles exhibit a SAM error ranging from $<10^\circ$ under clear conditions to $>20^\circ$ under severe environmental challenges. $\text{SAM}_\lambda$ and $\text{MAE}_\lambda$ are consistently lower than $\text{SAM}$ and $\text{MAE}$, confirming that heavily attenuated long-wavelength channels are the most challenging to recover. Although absolute spectral alignment remains out of reach due to the severe physical distortions introduced by the underwater medium, the recovered profiles capture meaningful coarse spectral trends, demonstrating the feasibility of the proposed framework.

Accuracy is governed primarily by water severity and channel count: performance degrades sharply in more turbid water (Jerlov~II) and coarsens as $C$ drops, confirming that hyperspectral resolution is essential for disambiguating the underdetermined inversion. Sensor noise, ranging errors and depth variations primarily cause degradation in the sensitive 600--700 nm range, reflected in the stability of $\text{SAM}_\lambda$ compared to the high sensitivity of standard $\text{SAM}$.

The ablation study validates the importance of each core component, as the independent removal of the observation filter, regularization term, and basis decomposition yields a significant performance drop. On downstream restoration benchmarks, our method clearly outperforms channel-wise baselines, validating that continuous spectral domain processing drives the improvement.

\subsection{Limitations and Future Work}
We identify physical and modeling limitations that outline opportunities for future work:
\begin{itemize}
    \item \textbf{Signal starvation:} At large depths or distant viewpoints, heavily attenuated long-wavelength bands (particularly red) fall into the sensor noise floor, leaving insufficient signal for recovery.
    \item \textbf{Ranging accuracy:} Like all physics-based restoration methods, ranging errors propagate through attenuation inversion. Future work should explore hardware-integrated ranging sensors or learned monocular/multi-view estimation.
    \item \textbf{Active lighting:} Our forward model assumes uniform ambient illumination. While active lighting (\eg, flashlights) could resolve signal starvation, its spatially varying intensity violates current assumptions and requires an extension of the forward model.
    \item \textbf{Non-Lambertian reflectance:} We assume Lambertian surfaces throughout. The water parameter estimation in the sparse recovery stage is robust to violations of this assumption, as it can suppress outlier observations. The dense stage operates independently per view, so for specular or non-Lambertian surfaces the recovered radiance is view-dependent and reflects appearance rather than intrinsic diffuse reflectance. Separating the diffuse component remains an open problem.
    \item \textbf{Real-world validation:} Developing the experimental setup required for real-world validation is necessary to fully characterize robustness against physical distractors, imaging challenges, and sensor artifacts.
    \item \textbf{Algorithmic refinement:} Moving beyond independent pixel-wise processing by integrating spatial or deep image priors, alongside the joint refinement of geometry and appearance, could significantly improve robustness to noise, distractors, and distance inaccuracies.
\end{itemize}

Overcoming these constraints will fully unlock the framework's operational utility across diverse marine environments. Quantitative tasks like infrastructure inspection or benthic mapping can directly exploit the recovered spectra, while the enhanced restoration could benefit detection pipelines and real-time vision enhancement for underwater vehicles or divers.

\section{Conclusion}
\label{sec:conclusion}

By introducing the first framework to recover continuous spectral radiance from underwater hyperspectral images, this work presented a stepping stone toward quantitative marine vision. Our pipeline performs robust sparse multi-view estimation of water parameters followed by efficient full-image recovery of scene radiance, each represented within a continuous spectral basis. Experiments on physics-based synthetic data demonstrate the feasibility of the proposed formulation and show that it can recover physically meaningful spectral signatures under a range of environmental and imaging conditions. We hope this work establishes spectral recovery as a foundation for future research on physically grounded underwater vision.

\bibliographystyle{abbrv}
\bibliography{main}

\clearpage
\appendix

\section{The Case for Spectral Recovery} \label{app:broadband}

To contextualize the advantages of our continuous spectral framework, we highlight the systematic errors introduced by conventional physics-based broadband (\eg, RGB) restoration models.

\paragraph{The monochromatic approximation.}
Under the assumption of a hypothetical monochromatic sensor with an infinitely narrow spectral sensitivity modeled as a Dirac delta function ($S_c(\lambda) = \delta(\lambda - \lambda_c)$), the image formation model in \cref{eq:pixel_value_full} collapses to:
\begin{equation} \label{eq:monochromatic_model}
    I_c = J_c\, e^{-\beta(\lambda_c)z} + B^\infty(\lambda_c)\bigl(1 - e^{-\beta(\lambda_c)z}\bigr).
\end{equation}
In this monochromatic space, $\beta(\lambda_c)$ and $B^\infty(\lambda_c)$ are exact, spatially invariant physical constants of the medium.

\paragraph{Broadband effective parameter errors.}
To bypass the intractable integrals of continuous broadband sensors, conventional frameworks typically copy this monochromatic shorthand, substituting continuous profiles with per-channel effective constants $\beta_c$ and $B^{\infty}_c$:
\begin{equation} \label{eq:simplified_model}
    I_c = J_c\,e^{-\beta_c z} + B^{\infty}_c\bigl(1 - e^{-\beta_c z}\bigr).
\end{equation}
Equating this mathematical simplification to the true physical integration model yields the definitions of these effective coefficients:
\begin{equation}
    \beta_c = -\frac{1}{z} \ln \left( \frac{\int_\Omega S_c(\lambda) L_0(\lambda) e^{-\beta(\lambda)z} \mathrm{d}\lambda}{\int_\Omega S_c(\lambda) L_0(\lambda) \mathrm{d}\lambda} \right).
\end{equation}
This mathematically demonstrates that the effective coefficient $\beta_c$ inherently conflates the sensor sensitivity $S_c(\lambda)$, the continuous water parameters $\beta(\lambda)$, the target scene distance $z$, and the unknown surface radiance $L_0(\lambda)$.

Akkaynak and Treibitz~\cite{akkaynak2018revised} partially mitigated this error by decoupling the direct and backscatter pathways:
\begin{equation} \label{eq:revised_model}
    I_c = J_c \,e^{-\beta_c^{(D)} z} + B^{\infty}_c\bigl(1 - e^{-\beta_c^{(B)} z}\bigr).
\end{equation}
While separating $\beta_c^{(D)}$ and $\beta_c^{(B)}$ yields higher accuracy, these variables remain proxy representations that inherently fluctuate with the scene's geometry and reflectance. Consequently, optimizing them as global constants introduces inevitable structural and chromatic artifacts, a limitation bypassed by lifting the restoration into the continuous spectral domain as described in \cref{sec:methods}.

\section{Methodological Details} \label{app:method_details}

\subsection{Choice of Spectral Bases}\label{app:spectral_bases}

Low-dimensional data-driven spectral bases have long been used to represent natural reflectance spectra, with studies showing that a small number of principal components capture most of the variation observed in real-world materials~\cite{cohen1964dependency}. Motivated by these results, we evaluated PCA bases for scene radiance learned from a standard reflectance dataset \cite{vrhel1994measurement}, as well as PCA bases for the water attenuation and veiling-light spectra learned from the oceanographic dataset containing spectral properties of the Jerlov water types~\cite{williamson2023depth}. We compared these representations against cubic B-spline bases and consistently found the latter to produce more stable optimization and superior reconstruction quality.

We hypothesize that this behavior is due to the global support of PCA basis functions. Updating a single coefficient alters the reconstructed spectrum across the entire wavelength range, coupling well-constrained and weakly constrained spectral regions. In our setting, this effect was particularly pronounced in the long-wavelength (red) region, where the measurements provide relatively limited information. As a result, optimization steps intended to improve the red region often introduced undesirable changes elsewhere in the spectrum. In contrast, the local support of cubic B-splines confines each update to a limited wavelength interval, allowing the optimizer to adjust different spectral regions more independently and leading to consistently more stable convergence.

These observations do not imply that PCA bases are fundamentally unsuitable for the problem. Rather, they suggest that the current optimization procedure does not adequately account for the strong coefficient coupling induced by globally supported basis functions. Developing optimization strategies that better exploit PCA bases remains an interesting direction for future work.

\subsection{Observation Filtering}\label{app:obs_filtering}

A set of multi-view observations $\left\{ (\mathbf{I}_{ij}, z_{ij}) \mid j \in \mathcal{V}_i \right\}$ corresponding to a 3D point $P_i$ is discarded under any of the following conditions:
\begin{enumerate}
    \item \textbf{Insufficient distance diversity:} Distance values $z_{ij}$ are partitioned into discrete intervals of width $\delta_z$ below a distance ceiling $\tau_z$:
    \begin{equation}
        \mathcal{B}_i = \left\{ \left\lfloor \frac{z_{ij}}{\delta_z} \right\rfloor \;\middle|\; j \in \mathcal{V}_i \;\wedge\; z_{ij} < \tau_z \right\}.
    \end{equation}
    The point is removed if $|\mathcal{B}_i| < \tau_{\text{bins}}$, ensuring an optimal distance baseline to mathematically constrain the non-linear medium attenuation curves.

    \item \textbf{High geometric or tracking uncertainty:} Let $\epsilon_i$ denote the geometric uncertainty of point $P_i$ (defined as the reprojection error in Structure-from-Motion frameworks, or the spatial covariance/depth uncertainty provided by a ranging sensor). The point is rejected if:
    \begin{equation}
        \epsilon_i > \tau_{\epsilon}.
    \end{equation}

    \item \textbf{High local image contrast:} To prevent sub-pixel boundary color blending and edge artifacts, the mean local contrast variance $\bar{s}_i$ is computed using the spatial standard deviation $\sigma_c$ over a $3 \times 3$ pixel window $\mathcal{W}(ij)$ across all channels $C$:
    \begin{equation}
        \bar{s}_i = \frac{1}{|\mathcal{V}_i| \cdot C} \sum_{j \in \mathcal{V}_i} \sum_{c=1}^{C} \sigma_c\bigl(\mathcal{W}(ij)\bigr).
    \end{equation}
    Points are rejected if $\bar{s}_i > \tau_s$.
\end{enumerate}

\subsection{Sparse Recovery}

\subsubsection{Dynamic Outlier Suppression}\label{app:outlier}

To suppress outliers caused by mismatched keypoints, noise, caustics, or non-Lambertian surfaces, we incorporate a dynamic, per-observation reweighting scheme during optimization.

Let $e_{ij}^{(t)}$ denote the root-mean-square error (RMSE) representing the modeling error between the observed pixel intensity $\mathbf{I}_{ij}$ and the forward model's prediction $\hat{\mathbf{I}}_{ij}(\boldsymbol{\Theta}^{(t)})$ for point $i$ in viewpoint $j$ at optimization iteration $t$, calculated across all channels $C$:
\begin{equation}
e_{ij}^{(t)} = \sqrt{\frac{1}{C} \sum_{c=1}^C \left( I_{ij,c} - \hat{I}_{ij,c}(\boldsymbol{\Theta}^{(t)}) \right)^2}.
\end{equation}
After an initial stabilization period ($t > 10$), we calculate a data-driven outlier boundary based on the global median performance of all active observations:
\begin{equation}
\tau_{\text{outlier}} = 2.5 \cdot \text{median}_{i,j} \big( e_{ij}^{(t)} \big).
\end{equation}
If the residual of a specific viewpoint measurement exceeds this adaptive threshold ($e_{ij}^{(t)} > \tau_{\text{outlier}}$), its individual fitting weight $\omega_{ij}$ in the sparse objective (\cref{eq:sparse_objective}) is independently penalized via a decay factor $\gamma \in (0, 1)$:
\begin{equation}
\omega_{ij}^{(t+1)} = \gamma \cdot \omega_{ij}^{(t)}.
\end{equation}
By operating on a per-viewpoint basis $(i,j)$ rather than filtering all observations of $P_i$, this progressive scheme isolates and suppresses transient artifacts while preserving valid radiometric data from alternative viewpoints. Consequently, corrupted data in a single view does not degrade the optimization of the entire 3D point.

\subsubsection{Schur Complement Matrix Decomposition}\label{app:schur}

The joint optimization problem presented in \cref{sec:methods_sparse} scales linearly with the number of observed points $N$. The regularized Levenberg-Marquardt update solves a linear system of normal equations $(\mathbf{H} + \mu \mathbf{I})\Delta \boldsymbol{\Theta} = -\mathbf{g}$. Due to the independent nature of the local point radiance weights $\mathbf{w}_i^{(L)}$, the Hessian matrix exhibits a sparse block-arrowhead configuration:
\begin{equation}
\begin{pmatrix}
\mathbf{A}_1 & & & \mathbf{B}_1 \\
& \ddots & & \vdots \\
& & \mathbf{A}_{N} & \mathbf{B}_{N} \\
\mathbf{B}_1^T & \dots & \mathbf{B}_{N}^T & \mathbf{C}
\end{pmatrix}
\begin{pmatrix} \Delta \mathbf{w}_1^{(L)} \\ \vdots \\ \Delta \mathbf{w}_{N}^{(L)} \\ \Delta \mathbf{w}_{\text{global}} \end{pmatrix} =
\begin{pmatrix} \mathbf{\mathbf{r}}_1 \\ \vdots \\ \mathbf{\mathbf{r}}_{N} \\ \mathbf{\mathbf{r}}_{\text{global}} \end{pmatrix},
\end{equation}
where $\mathbf{A}_i$ are block-diagonal matrices corresponding to local points, and $\mathbf{C}$ tracks global medium updates $\mathbf{w}^{(\beta)}, \mathbf{w}^{(B)}$. We compute the Schur complement to marginalize out the local parameters, reducing the central inversion task to a small, dense system solely bounding the global medium weights:
\begin{equation}
\left( \mathbf{C} - \sum_{i=1}^N \mathbf{B}_i^T \mathbf{A}_i^{-1} \mathbf{B}_i \right) \Delta \mathbf{w}_{\text{global}} = \mathbf{\mathbf{r}}_{\text{global}} - \sum_{i=1}^N \mathbf{B}_i^T \mathbf{A}_i^{-1} \mathbf{\mathbf{r}}_i.
\end{equation}
The local parameters are subsequently updated in parallel via back-substitution.

\subsection{Dense Recovery}

\subsubsection{Derivation of the Channel-Space Transmission Approximation}\label{app:transmission_derivation}

\begin{figure}[ht]
    \centering
    \includegraphics[width=0.7\linewidth]{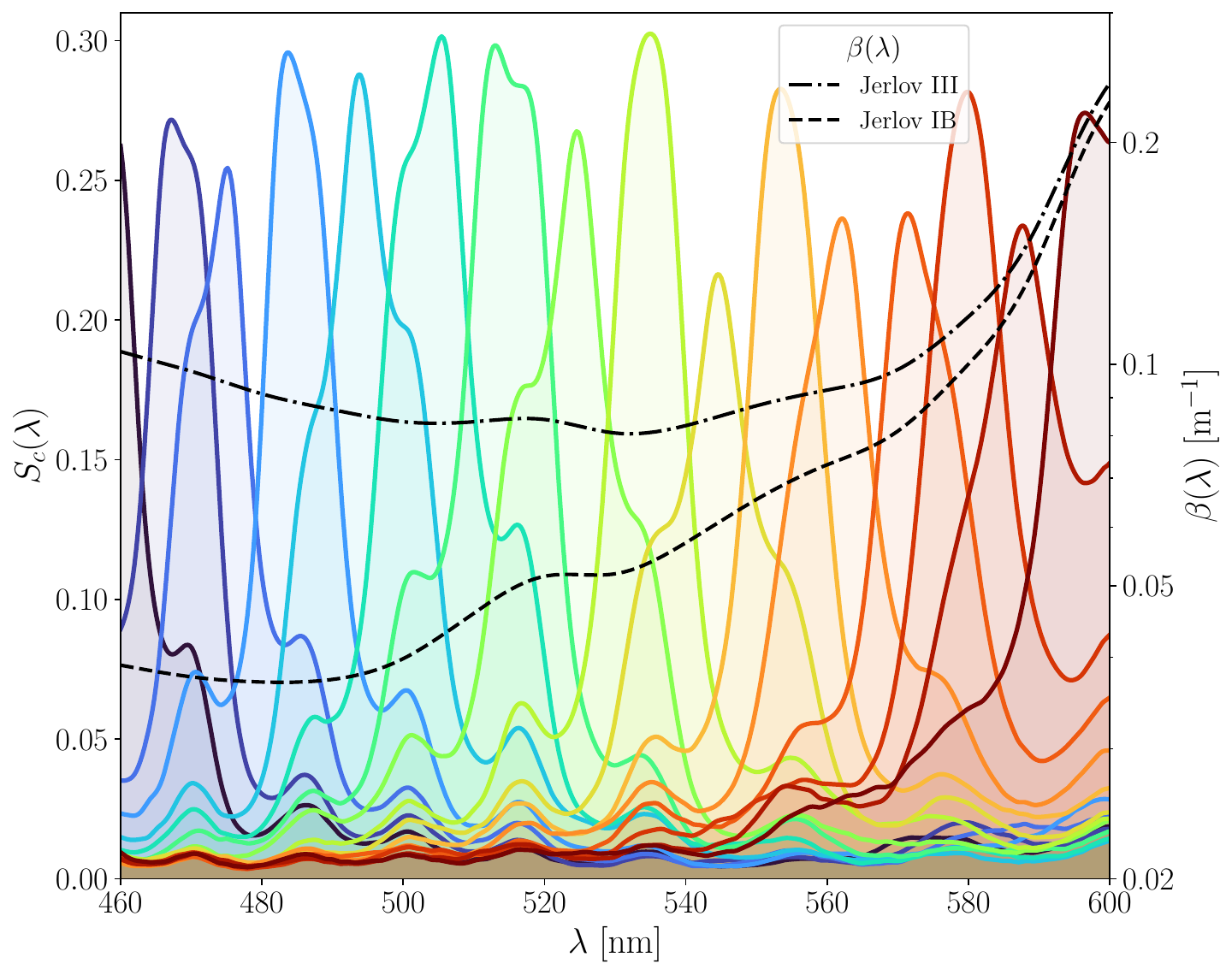}
    \caption{Spectral sensitivity profiles of a real HSI camera (\textit{imec snapshot VIS} \cite{imecsnapshotvis}) with $C=16$ and $\Omega_c=[460\text{ nm}, 600 \text{ nm}]$.}
    \label{fig:spectral_response_imec}
\end{figure}

To arrive at the approximation $\hat{I}_c^{(D)} \approx J_c \hat{T}_c$ from \cref{I_D_hat}, we assume that the continuous medium attenuation $\hat{\beta}(\lambda)$ varies minimally within the narrow bandwidth $\Omega_c$ of channel $c$. As illustrated in \cref{fig:spectral_response_imec}, this is a good approximation for HSI cameras. This allows us to approximate it with a constant effective channel coefficient, $\hat{\beta}(\lambda) \approx \beta_c$ for all $\lambda \in \Omega_c$.

The estimated direct signal component is physically modeled as:
\begin{equation} \label{eq:id_continuous}
\hat{I}_c^{(D)} = \int_{\Omega_c} S_c(\lambda) L_0(\lambda) e^{-\hat{\beta}(\lambda)z} \,\mathrm{d}\lambda.
\end{equation}
Pulling the constant narrowband attenuation term $e^{-\beta_c z}$ out of the integral yields:
\begin{equation} \label{eq:id_pulled}
\hat{I}_c^{(D)} \approx e^{-\beta_c z} \int_{\Omega_c} S_c(\lambda) L_0(\lambda) \,\mathrm{d}\lambda = J_c e^{-\beta_c z},
\end{equation}
where $J_c = \int_{\Omega_c} S_c(\lambda) L_0(\lambda) \,\mathrm{d}\lambda$ represents the ideal latent channel intensity at $z=0$.

By defining the effective channel transmissivity directly as $\hat{T}_c = e^{-\beta_c z}$, \cref{eq:id_pulled} immediately simplifies to:
\begin{equation}
\hat{I}_c^{(D)} \approx J_c \hat{T}_c.
\end{equation}
To ensure this discrete channel-space representation aligns with continuous physics, $\hat{T}_c$ is equated to the continuous transmission integrated across the normalized channel sensitivity ($\int_{\Omega_c} S_c(\lambda)\,\mathrm{d}\lambda = 1$):
\begin{equation} \label{eq:tc_continuous}
\hat{T}_c = e^{-\beta_c z} \equiv \int_{\Omega_c} S_c(\lambda) e^{-\hat{\beta}(\lambda)z} \,\mathrm{d}\lambda.
\end{equation}

\subsubsection{Adaptive Regularization Mechanics for Spectral Lifting}\label{app:adaptive_reg}

The regularization weight $\xi$ in \cref{eq:dense_pixel_ridge_problem} adaptively balances sensor noise against numerical stability on a per-pixel basis. To achieve this, we compute a dynamic, channel-wise noise-to-signal score $\eta$ and map it to $\xi$ using:
\begin{equation}\label{eq:local_signal_quality}
\eta = \max_c \left(\frac{\sqrt{\sigma^2}}{\sqrt{\hat J_c^2 + \sigma^2}}\right), \quad \xi = \min\!\left( \xi_0 \!\left(\frac{\eta}{1 - \eta}\right)^2,\; \xi_{\max}\right),
\end{equation}
where $\sigma^2$ is the sensor's noise variance.

Physically, this formulation prevents unconstrained pseudo-inverses from amplifying noise in highly attenuated color channels. For example, in deep water columns where red light degrades rapidly, a standard inversion would overcompensate for the weak signal, resulting in severe red-bleeding or false-color artifacts. By taking the maximum across all channels ($\max_c$), we regularize based on the worst-case conditioning. This joint damping protects the numerical system from its weakest link while preserving inter-channel ratios (chromaticity), which would otherwise be distorted by independent channel regularizers.

\section{Experimental Details} \label{app:experiment_details}

\subsection{Synthetic Data Generation} \label{app:synth_data_gen}

We use Mitsuba 3~\cite{jakob2022mitsuba3} to simulate wavelength-dependent underwater light transport by numerically solving the radiative transfer and volume rendering equations. Materials are specified via continuous reflectance/emission spectra, water via wavelength-dependent absorption and scattering coefficients, and camera sensors via spectral sensitivity curves. This yields low-level physical simulation of multi-scattering, forward-scattering blur, depth-dependent absorption, and wave caustics.

\begin{figure}[ht]
    \centering
    \begin{subfigure}[b]{0.48\linewidth}
        \centering
        \includegraphics[width=\linewidth]{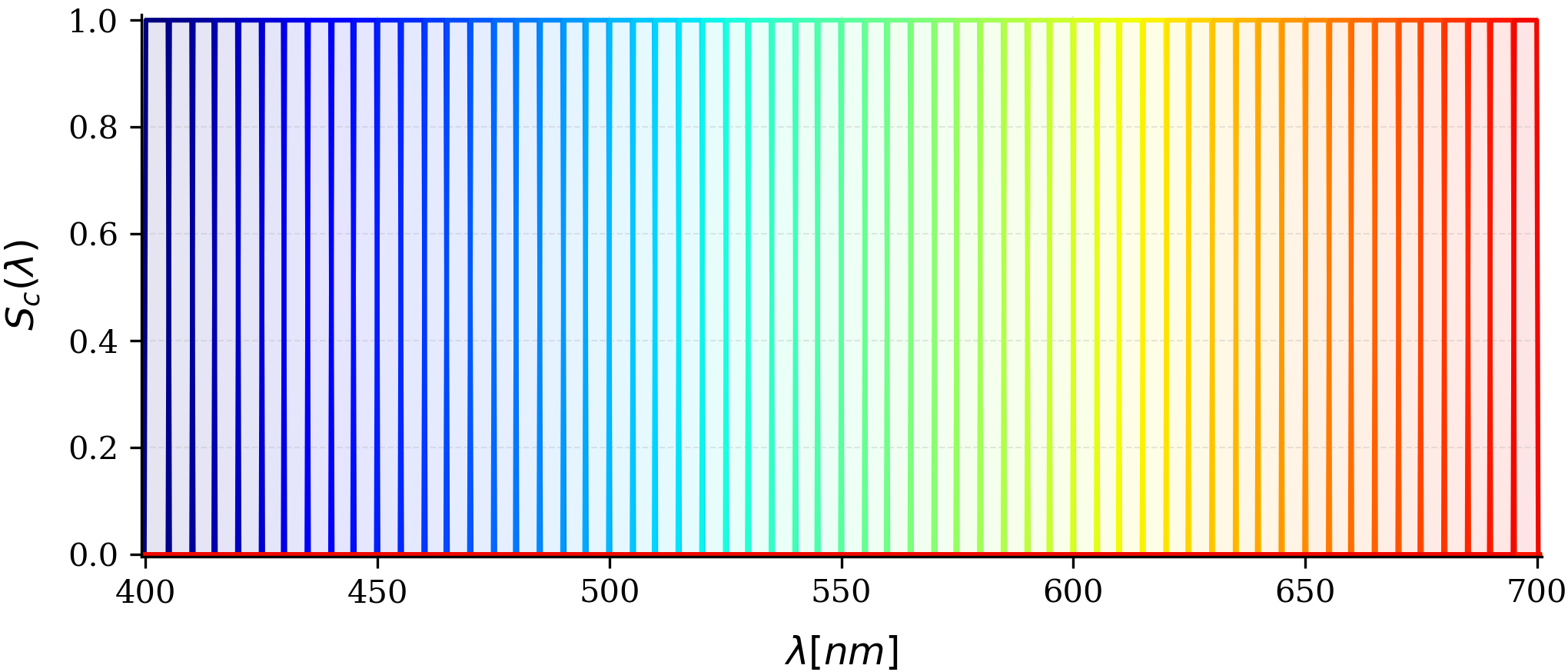}
        \caption{Virtual sensor}
        \label{fig:sens_curves_gt}
    \end{subfigure}
    \hfill
    \begin{subfigure}[b]{0.48\linewidth}
        \centering
        \includegraphics[width=\linewidth]{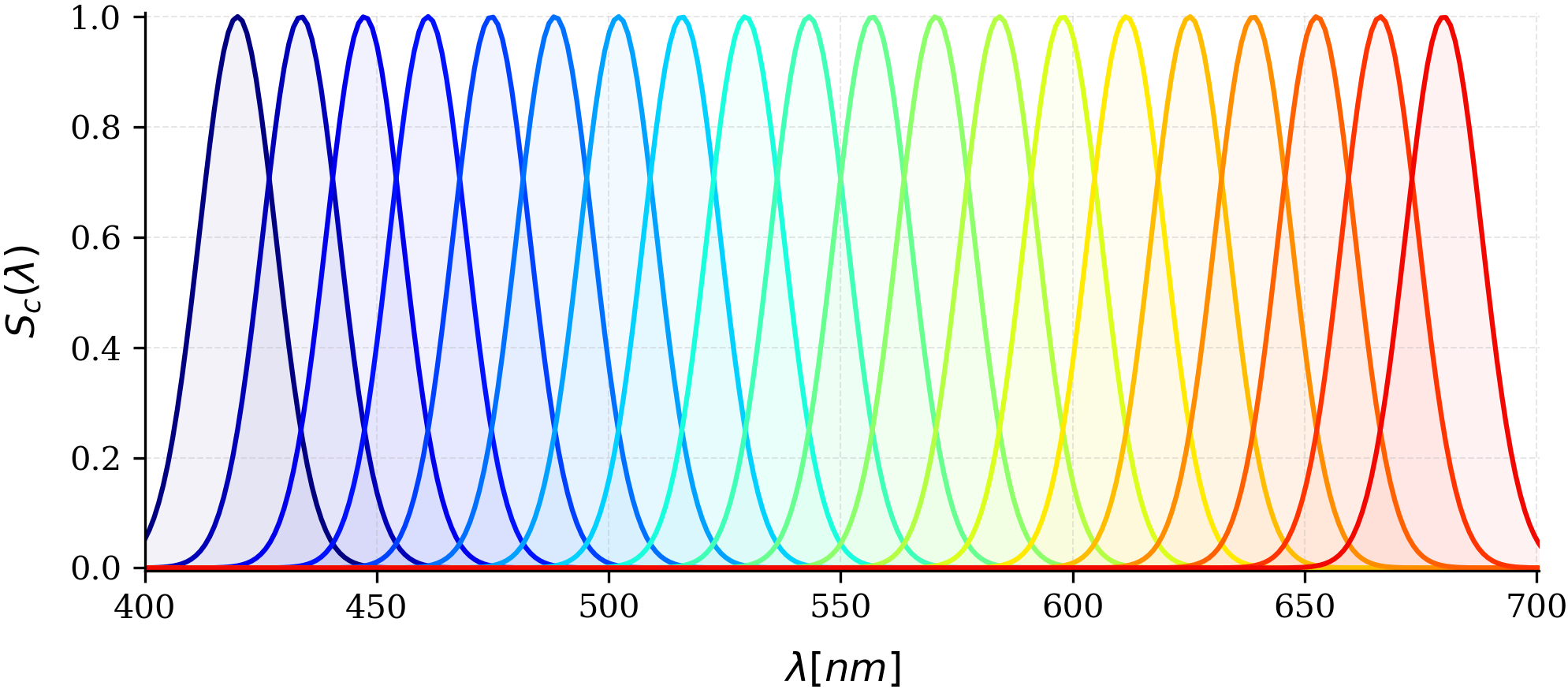}
        \caption{Synthetic HSI sensor ($C=20$)}
        \label{fig:sens_curves_gaussian}
    \end{subfigure}
    \caption{\textbf{Synthetic sensor sensitivity functions $S_c(\lambda)$.} (a) The idealized, non-overlapping rectangular filter functions used to discretize the continuous spectrum into ``ground-truth'' radiance profiles. (b) The synthesized hyperspectral camera sensitivities modeled as a bank of overlapping Gaussians spanning the visible spectrum with an overlap factor of 0.6.}
    \label{fig:synth_hsi_sensitivities}
\end{figure}

Our Python wrapper (to be open-sourced upon publication) supports multi-view rendering of dynamic scenes (waves, marine snow), holds a sensor/water asset library (both real-world and synthetic), and computes discretized ground-truth radiance spectra using a virtual sensor composed of contiguous bands of $\Delta \lambda=5$ nm wide (see \cref{fig:sens_curves_gt}). All images are rendered at a spatial resolution of $(W, H) = (480, 320)$ pixels.

To accurately resolve complex media interactions, scenes are rendered using a volumetric path tracer equipped with spectral multiple importance sampling (MIS)~\cite{miller2019null}. This approach jointly samples both phase function directional scattering and light source geometries, minimizing variance in areas governed by dense forward-scattering and caustic light concentrations.

In Mitsuba~3, homogeneous media are characterized by their spectral \textit{single-scattering albedo}, spectral \textit{extinction coefficient}, and directional \textit{phase function}. The albedo and extinction coefficients for standard water types are sourced from open oceanographic databases, as detailed below. The phase function, dictating the angular distribution of light scattered by particles, is implemented using the Henyey--Greenstein formula \cite{henyey1941diffuse} with an anisotropy factor of $g=0.924$, representing the characteristic forward-scattering profile of natural oceanic particles \cite{mobley1994light}.

We generate evaluation data by creating and exporting scenes from Blender. Because Blender assets typically use RGB colors rather than continuous wavelengths, Mitsuba~3's spectral engine automatically upsamples these base RGB material definitions into realistic, continuous spectral power distributions using the spectral upsampling procedure introduced by Jakob and Hanika~\cite{jakob2019low}. To simulate experimental HSI acquisition, the continuous radiance fields are integrated against custom sensor sensitivity curves $S_c(\lambda)$, modeled as a bank of $C$ overlapping Gaussians spanning the $[400\text{ nm}, 600\text{ nm}]$ spectrum with an overlap factor of 0.6. Specifically, the overlap factor is defined as the ratio of the Gaussian standard deviation to the peak-to-peak spacing $\Delta_c$ between adjacent channel centers, such that $\sigma = 0.6 \times \Delta_c$. \Cref{fig:sens_curves_gaussian} illustrates these profiles for $C=20$.

\subsection{Collecting Multi-View Observations}\label{app:mv_collection}

To collect multi-view observations from the synthetic data, we back-project the rendered distance maps into a global 3D point cloud using known camera intrinsics and extrinsic poses. We randomly subsample these 3D points and reproject them to the viewpoints that survive a $z$-buffer occlusion check. To simulate the limitations of underwater tracking, where visibility degrades over distance, the reprojection to these occlusion-free views is executed probabilistically, with a selection probability defined as $e^{-\psi z}$, using $\psi = 0.2\,\text{m}^{-1}$ throughout our experiments. For a valid projection of $P_i$ into view $j$, we extract the local pixel channel intensities via bilinear interpolation to form the observation tuple $(\mathbf{I}_{ij}, z_{ij})$.

\subsection{Scene Details}\label{app:scenes}

\begin{figure}[ht]
    \centering
    \includegraphics[width=0.99\linewidth]{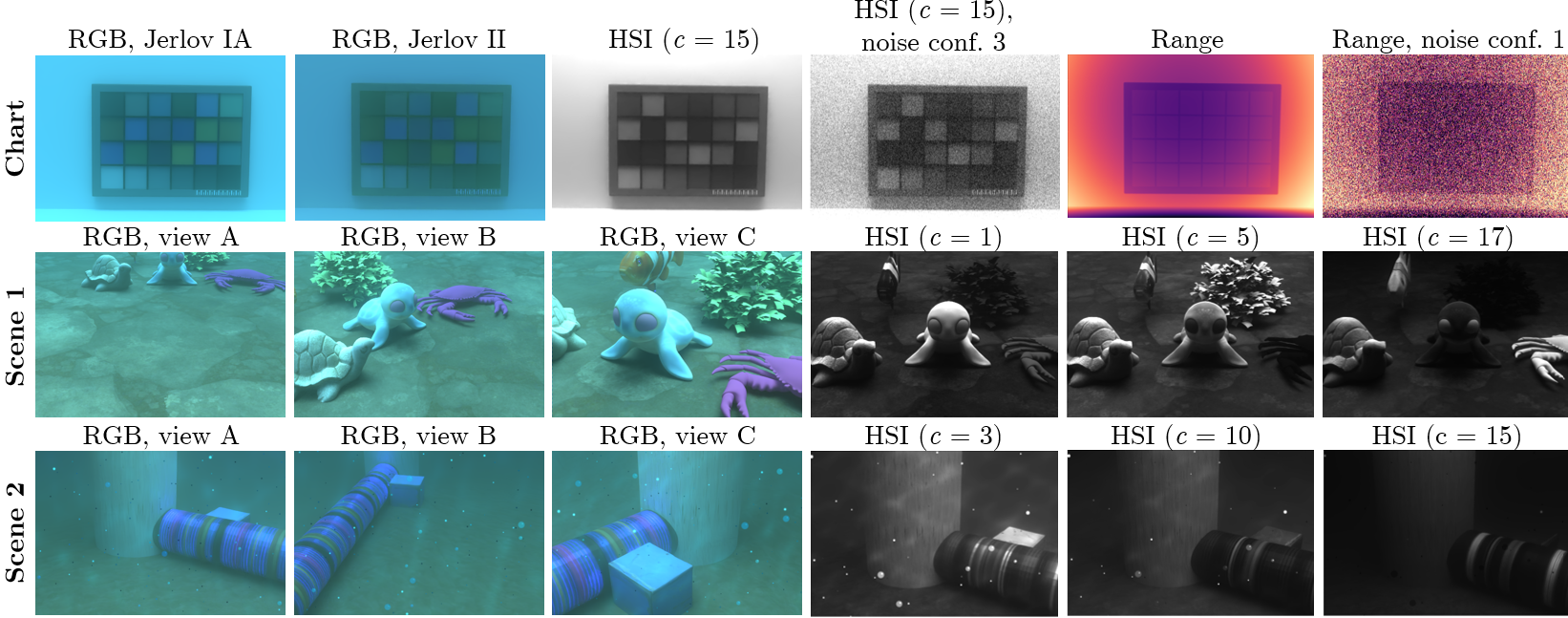}
    \caption{\textbf{Illustrative examples of the evaluation scenes.} \textbf{Top:} The \textit{Chart} scene rendered in RGB across two water types, alongside HSI views (single channel shown with/without noise) and distance maps (with/without noise). \textbf{Middle/Bottom:} \textit{Scene 1} \& \textit{2} rendered in RGB from three perspectives, followed by three distinct channels of a HSI render. The RGB views are brightened for visualization purposes.}
    \label{fig:scene_previews}
\end{figure}

We evaluate our method on three simulated scenes: \textit{Chart}, \textit{Scene 1}, and \textit{Scene 2}, all illuminated using the Hosek--Wilkie sky-dome model~\cite{hosek2012analytic}. \Cref{fig:scene_previews} displays these scenes across various viewpoints, modalities and degradations.

\paragraph{Chart.} Used for accuracy and sensitivity analyses, this scene features a 24-color calibration chart against a uniform gray backdrop. The camera linearly approaches the chart from $20\,\text{m}$ to $2\,\text{m}$ over $12$ frames.

\paragraph{Scene 1.} Features five colorful marine assets (\eg, turtles, a crab, a fish) on a stone-paved floor, captured over $15$ frames along a $10\,\text{m}$ trajectory. The environment simulates clear water (Jerlov IA) at a $5\,\text{m}$ depth.

\paragraph{Scene 2.} Presents a challenging environment at a $10\,\text{m}$ depth in more attenuating water (Jerlov IB). It contains a white pillar, a blue box, and colored cables on a sandy floor, overlaid with three types of procedural dynamic distractors: \textit{Marine snow} is generated as 3D localized intensity blobs with a heavy-tailed power distribution for specular highlights, where near-lens particles are additively composited to simulate bokeh blur. \textit{Air bubbles} are modeled as soft refractive rings featuring an ambient-filled core and an offset, high-contrast specular reflection. \textit{Wave caustics} are generated via intersecting randomized sinusoidal wave fields, dynamically scaled by channel-specific Beer--Lambert absorption and smoothed via a distance-dependent Gaussian filter to simulate depth-induced high-frequency focus loss.

\subsection{Hyperparameter Configurations}\label{app:hyperparams}

To ensure reproducibility, \cref{tab:supp_hyperparameters} outlines the hyperparameter values utilized across all experiments, unless stated otherwise.

\begin{table}[ht]
\centering
\caption{Hyperparameter values.}
\label{tab:supp_hyperparameters}
\begin{tabular}{lcc}
    \toprule
    \textbf{Hyperparameter} & \textbf{Symbol} & \textbf{Value} \\
    \midrule

    \addlinespace
    \multicolumn{3}{l}{\textbf{Spectral bases}} \\
    Radiance basis dimensionality & $K_L$ & 10 \\
    Beam attenuation basis dimensionality & $K_\beta$ & 8 \\
    Veiling light basis dimensionality & $K_B$ & 8 \\

    \addlinespace
    \multicolumn{3}{l}{\textbf{Filtering}} \\
    Min. distance bins & $\tau_\text{bins}$ & 3 \\
    Distance bin width & $\delta_z$ & 0.25 m \\
    Max. reliable distance & $\tau_z$ & 3 m \\
    Max. reprojection error & $\tau_\epsilon$ & 1.5 pix \\
    Max. contrast & $\tau_s$ & 0.02 \\

    \addlinespace
    \multicolumn{3}{l}{\textbf{Regularization}} \\
    Huber delta & $\delta$ & $0.001$ \\
    Image noise variance & $\sigma^2$ & $0.001$ \\
    Outlier weight decay factor & $\gamma$ & 0.98 \\
    Base spectral regularization weight & $\xi_0$ & $0.01$ \\
    Max. spectral regularization weight & $\xi_{\max}$ & $0.001$ \\

    \bottomrule
\end{tabular}
\end{table}

\subsection{Sensitivity Study Configurations} \label{app:sensitivity}

This section details the parameterizations and noise configurations swept during the sensitivity analysis presented in \cref{sec:sensitivity}. The depth variation simply refers to varying subsurface scene depths, and the variation of the number of sensor channels ($C$) controls the set of sensitivity functions, generated as described in \cref{app:synth_data_gen}. The other variables are defined below.

\paragraph{Water type.}
The optical properties of the medium heavily influence spectral attenuation. In our sensitivity analysis, we evaluate performance across three distinct water profiles:
\begin{itemize}
    \item \textbf{Jerlov IA:} A standard optical water type representing clear open-ocean water with minimal attenuation, serving as our baseline.
    \item \textbf{Jerlov II:} A more turbid coastal/oceanic water type exhibiting substantially higher attenuation, particularly in the shorter blue wavelengths.
    \item \textbf{Clear:} A custom-defined synthetic profile simulating clear freshwater environments (\eg, highly transparent alpine lakes) characterized by low overall attenuation across the visible spectrum.
\end{itemize}
The Jerlov water coefficients $\beta(\lambda)$ are sourced from the reference database compiled by Williamson and Hollins \cite{williamson2023depth}.

\paragraph{Image noise.}
Sensor noise is simulated using a comprehensive hybrid model combining zero-mean additive Gaussian noise and signal-dependent shot noise. The configurations are parameterized by their respective intensity fractions ($\gamma_{\text{shot}}$ and $\gamma_{\text{gauss}}$) relative to the mean image intensity:
\begin{itemize}
    \item \textbf{Conf. 1:} Low Gaussian, high shot noise ($\gamma_{\text{shot}} = 0.03$, $\gamma_{\text{gauss}} = 0.001$).
    \item \textbf{Conf. 2:} High Gaussian, low shot noise ($\gamma_{\text{shot}} = 0.01$, $\gamma_{\text{gauss}} = 0.003$).
    \item \textbf{Conf. 3:} High Gaussian, high shot noise ($\gamma_{\text{shot}} = 0.03$, $\gamma_{\text{gauss}} = 0.003$).
\end{itemize}
\Cref{fig:scene_previews} shows a single channel of a HSI image corrupted by Conf.~3 noise.

\paragraph{Distance error.}
To simulate realistic depth-sensor uncertainty and ranging failures, we introduce pixelwise distance errors via a quadratic Gaussian noise model (where the standard deviation $\sigma_z$ scales quadratically with true distance $z$, $\sigma_z = \sigma_0 \cdot z^2$) coupled with salt-and-pepper missing value outliers.
\begin{itemize}
    \item \textbf{Conf. 1:} Base error standard deviation $\sigma_0 = 0.01$, outlier pixel fraction = $0.01$.
    \item \textbf{Conf. 2:} Base error standard deviation $\sigma_0 = 0.03$, outlier pixel fraction = $0.01$.
    \item \textbf{Conf. 3:} Base error standard deviation $\sigma_0 = 0.01$, outlier pixel fraction = $0.03$.
\end{itemize}
\Cref{fig:scene_previews} displays a distance map corrupted by Conf.~1 noise.

\paragraph{Distance diversity.}
Distance diversity governs the density and spatial distribution of the views along the optical axis, testing the optimization's sensitivity to multi-view geometric constraints:
\begin{itemize}
    \item \textbf{Conf. 1 (uniform downsampling):} The trajectory is uniformly downsampled by a factor of 2 (omitting every odd frame), resulting in $6$ views spanning the full $2\text{--}20\,\text{m}$ range.
    \item \textbf{Conf. 2 (truncated range):} The extreme near and far viewpoints are truncated, restricting the optimization strictly to the middle portion of the trajectory ($6\text{--}14\,\text{m}$) across $6$ views.
\end{itemize}

\section{Image Restoration Experiments} \label{app:img_restoration}

Here, we expand on our image restoration experiments, first detailing the baseline methods before presenting a comparative analysis of the qualitative results.

\paragraph{Ours$^\dagger$.}
The channel-wise variant of our method follows the same overall pipeline of multi-view observation collection, filtering, and sparse-to-dense reconstruction, but replaces the continuous spectral representation of the water parameters with channel-wise effective (band-averaged) parameters, using \cref{eq:revised_model} as the forward model. Although this approximation is less physically faithful, it avoids solving the computationally demanding inverse problem in the continuous spectral domain for every pixel. During the sparse stage, the attenuation coefficients $\beta_c^{(D)}$, backscatter coefficients $\beta_c^{(B)}$, and background radiance $B^\infty_c$ are for each channel $c \in {1, \dots, C}$, eliminating the need for basis-function projections. During the dense stage, by operating directly in the channel domain, this variant eliminates the spectral lifting stage altogether.

\paragraph{Sea-Thru (extended).}
As a second baseline, we extended the single-image Sea-Thru~\cite{akkaynak2019sea} algorithm to accommodate arbitrary channel counts ($C$). The pipeline performs independent channel-wise estimation in three steps. First, backscatter is isolated by clustering the scene's distance map into evenly spaced bins and selecting the darkest pixels (bottom 1st percentile) within each bin to fit the asymptotic background radiance $B^\infty_c$ and backscatter coefficient $\beta^B_c$ via non-linear curve fitting. Second, the local wideband illuminant map $\hat{E}_c$ is estimated from the backscatter-removed image using a depth-aware Local Space Average Color (LSAC) iteration or a fast spatial Gaussian-blur approximation. Third, a range-dependent attenuation coefficient $\beta^{(D)}_c(z)$ is computed per pixel from the illuminant map and subsequently smoothed by fitting a spatially regularized two-term exponential model.

\paragraph{Note on baseline selection.}
While learning-based underwater restoration methods exist, they are predominantly designed for standard RGB images, lack trivial extensions to arbitrary channel counts, or treat image restoration as a black-box image-to-image translation task. By comparing against \textit{Ours}$^\dagger$ and \textit{Sea-Thru (extended)}, we isolate the exact performance gains attributable to our continuous spectral representation and multi-view optimization framework under identical physical assumptions.

\paragraph{Comparison.}
For a fair comparison, the proposed and baseline methods all use the same white balancing approach. Specifically, for each restored image, we compute the color correction matrix that optimally maps the restored image to the clear-air reference image, where optimality is defined in the mean absolute error (MAE) sense.

\begin{figure}[htb]
    \centering
    \includegraphics[width=0.99\linewidth]{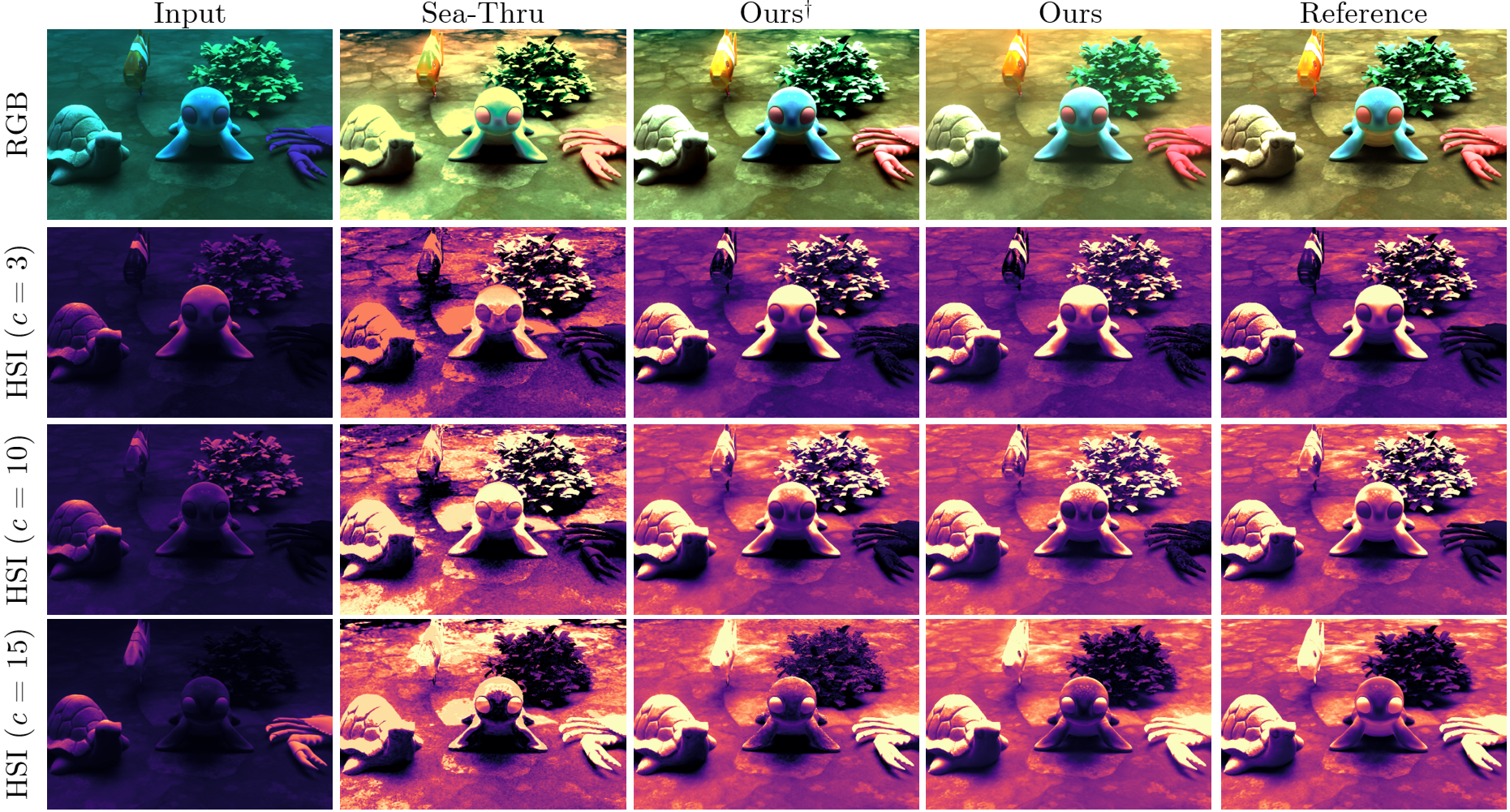}
    \caption{\textbf{Qualitative image restoration comparison on \textit{Scene 1}.} We compare our method against baseline approaches and the ground truth, visualizing both the RGB restorations and the HSI restorations across three selected channels.}
    \label{fig:restoration_scene1}
\end{figure}

As illustrated in \cref{fig:restoration_scene1}, compared to our approach, the baseline methods exhibit noticeable color shifts and significant degradation toward the background of the scene. This highlights a fundamental limitation: channel-effective parameters conflate optical and distance properties, making them inseparable in scenes with significant distance variation.

\begin{figure}[htb]
    \centering
    \includegraphics[width=0.99\linewidth]{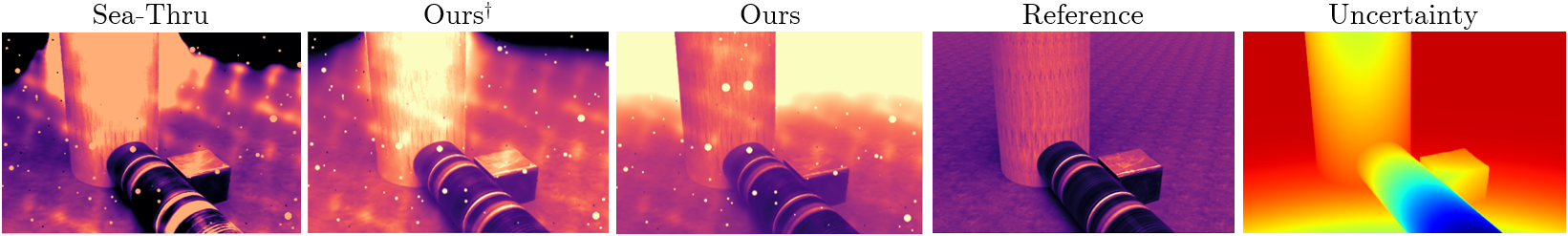}
    \caption{\textbf{Qualitative comparison under severe degradation and distractors on \textit{Scene 2}.} The figure shows channel 10 of the HSI restorations alongside the corresponding color-coded uncertainty map representing the local signal quality indicator $\xi$.}
    \label{fig:restoration_scene2}
\end{figure}

Conversely, \cref{fig:restoration_scene2} highlights the remaining limitations of the proposed approach. All evaluated methods struggle under the severe signal degradation caused by water attenuation, and the scene further demonstrates the impact of visual distractors. The rightmost panel displays an uncertainty map, computed as the pixel-wise local signal quality indicator $\xi$ (\cref{eq:local_signal_quality}). Evaluating our method using the Weighted Mean Absolute Error (WMAE), where the weights are defined as the inverse of the uncertainty, yields a $2.7\times$ reduction in error relative to the unweighted MAE. This demonstrates that our model is aware of its own predictive limitations, providing a reliable uncertainty map that can be directly utilized by downstream tasks.

These limitations highlight compelling avenues for future work. Beyond employing active lighting to mitigate severe signal attenuation, several architectural enhancements are possible. Specifically, our current dense recovery stage processes pixels independently. Integrating image priors (either traditional spatial smoothness constraints or deep learning-based priors) could guide the restoration process, facilitating a transition from pixel-wise estimation to patch- or image-level restoration. Additionally, explicit noise- and distractor-removal mechanisms could be integrated into the pipeline to improve robustness in complex environments.

\section{Practical Extensions} \label{app:practical}

\subsection{White Balancing}\label{app:white_balancing}

As described in \cref{subsec:metrics}, a fair quantitative evaluation requires remapping the recovered radiances to match the ambient illumination conditions of the ground-truth data. This step corrects for the depth-dependent chromatic bias introduced by underwater downwelling light. In our evaluation pipeline, this is achieved by computing and applying either an optimal illumination profile to map recovered radiances directly to the reference spectra (for continuous spectral recovery), or a color correction matrix to map the latent image to the reference target (for downstream image restoration).

In practical deployment scenarios, however, such reference images are unavailable. To achieve proper color constancy without a ground truth target, several alternative white balancing and relighting strategies can be employed:

\begin{itemize}
    \item \textbf{Gray-World Heuristic:} Assumes the average reflectance of the scene is isotropic and neutral gray, scaling individual channels uniformly to match a global average intensity.
    \item \textbf{Calibration Target:} Relies on a physical color checker captured both above the water surface and at the specific scene depth to derive an empirical illumination mapping function. If the absolute spectral reflectance of the target is known, this mapping can be directly inverted to recover the true surface reflectance $\rho(\lambda)$.
    \item \textbf{Analytical Relighting:} Computes the diffuse attenuation coefficient $K_d(\lambda)$ analytically from optimized inherent optical properties (IOPs) in closed-form following Haltrin's radiative transfer formulation \cite{haltrin1998apparent}. Given a scalar backscattering asymmetry fraction estimate $\tilde{b}_b$, the coefficient $K_d(\lambda)$ is approximated as:
    \begin{equation}
        K_d(\lambda) \approx \frac{\beta(\lambda)\,\tilde{b}_b}{\tilde{b}_b\bigl(1-B^\infty(\lambda)\bigr) + B^\infty(\lambda)}.
    \end{equation}
    The restored atmospheric radiance is then reconstructed via:
    \begin{equation}
        \hat L_0^{(\mathrm{atm})}(\lambda) = e^{K_d(\lambda)d_0} L_0(\lambda),
    \end{equation}
    or a regularized version (\eg, Tikhonov filtering) to maintain numerical stability at large operational depths. The backscattering probability parameter $\tilde{b}_b$ relates directly to the medium's phase function \cite{haltrin2002one}; for a typical oceanic Henyey--Greenstein distribution \cite{henyey1941diffuse} with an anisotropy factor of $g=0.924$, this fraction evaluates to $\tilde{b}_b \approx 0.0155$.

    \item \textbf{Radiometer Measurement:} Directly measures the atmospheric downwelling irradiance spectrum $E(0, \lambda)$ at the surface alongside the sub-surface downwelling irradiance $E(d_0, \lambda)$ at scene depth $d_0$. This enables the empirical calculation of the water column's spectral transmission profile, which is subsequently inverted to restore both the original scene radiance and surface reflectance.
\end{itemize}

\begin{figure}[ht]
    \centering
    \includegraphics[width=0.99\linewidth]{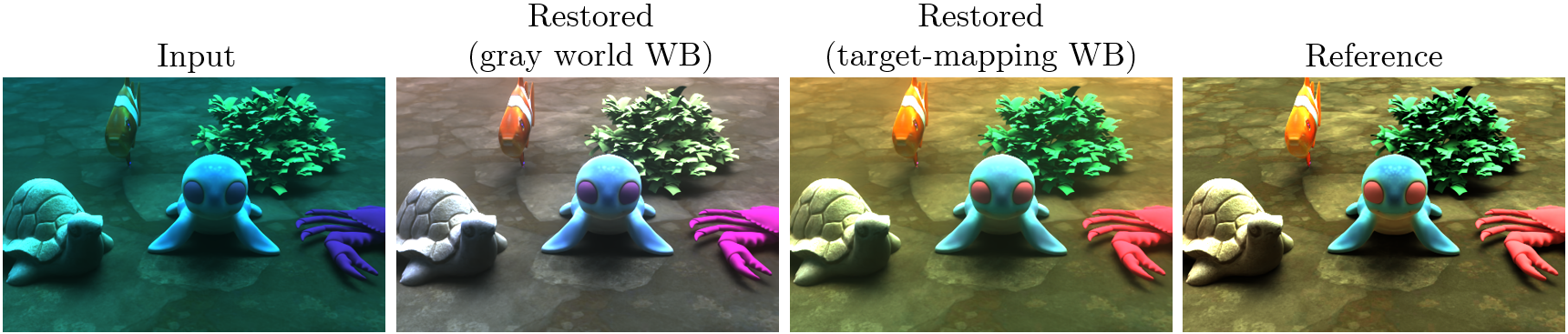}
    \caption{\textbf{Illustration of two different white balancing approaches.} (a) The original underwater image; (b) the restored image when applying the gray world heuristic in the white balancing stage; (c) the restored image when applying the optimal color correction matrix mapping the latent image to the reference image (as used in our experiments); (d) the reference image.}
    \label{fig:wb_comparison}
\end{figure}

\Cref{fig:wb_comparison} qualitatively illustrates the restored outputs obtained using the gray-world heuristic compared against the reference-targeted color mapping technique utilized in our evaluation experiments.

\subsection{Vision-Based Ranging}\label{app:ranging}

\begin{figure}[ht]
    \centering
    \includegraphics[width=0.99\linewidth]{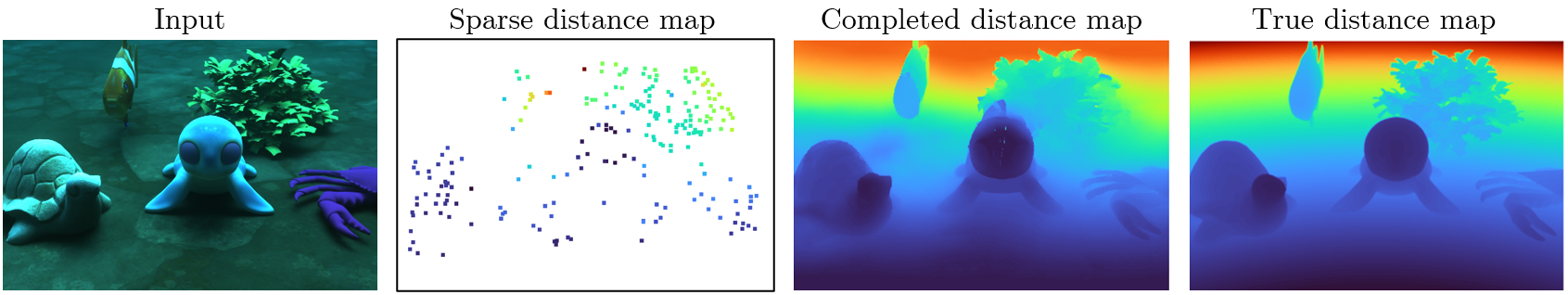}
    \caption{\textbf{Overview of the dense distance map estimation pipeline.} (a) Input underwater image; (b) sparse depth map derived from SfM; (c) estimated dense distance map via deep depth completion; (d) ground-truth dense distance map.}
    \label{fig:depth_completion}
\end{figure}

In practical deployment scenarios, ground-truth distance maps are typically unavailable and must be estimated via stereo vision, active ranging (\eg, sonar or LiDAR), or Structure-from-Motion (SfM). To illustrate this workflow, we consider a vision-based pipeline that combines SfM with deep depth completion to generate dense distance maps through the following steps, which are illustrated in \cref{fig:depth_completion}:

\begin{enumerate}
    \item \textbf{Sparse 3D reconstruction:} We execute COLMAP~\cite{schonberger2016structure} on the input image sequence to recover a sparse 3D point cloud, camera poses, and the camera intrinsic matrix $\mathbf{K}$. By projecting the reconstructed 3D points onto the image planes of their corresponding camera frames, we obtain a set of sparse depth maps.

    \item \textbf{Depth completion:} We utilize Marigold-DC~\cite{viola2025marigold}, a zero-shot depth completion network, to predict dense, structurally consistent depth maps $d(x,y)$ for each frame. Let $(c_x, c_y)$ denote the principal point and $(f_x, f_y)$ signify the focal lengths extracted from $\mathbf{K}$. To convert the depth maps into distance maps, each pixel-wise depth value $d(x,y)$ is converted into a distance value $z(x,y)$ via:
    \begin{equation}
        z(x,y) = d(x,y) \cdot \sqrt{1 + \left(\frac{x - c_x}{f_x}\right)^2 + \left(\frac{y - c_y}{f_y}\right)^2}.
    \end{equation}
\end{enumerate}

\Cref{tab:dist_comparison} presents an initial ablation of reconstruction errors when replacing idealized geometry with this automated tracking and depth estimation pipeline. These preliminary results suggest that our optimization framework can tolerate realistic tracking and depth artifacts without catastrophic degradation. Interestingly, replacing uniform ground-truth sampling with COLMAP-derived sparse points yields lower MAE in certain configurations. We hypothesize that feature-based SfM inherently selects keypoints that provide well-constrained optimization cues. Conversely, random ground-truth sampling inevitably includes points in dark regions where the subsequent optimization is poorly constrained, leading to higher overall reconstruction errors.

\begin{table}[ht]
\centering
\caption{Ablation of reconstruction accuracy across different combinations of ground-truth (True) and estimated (COLMAP / Marigold-DC) geometry on \textit{Scene 1}.}
\label{tab:dist_comparison}
\begin{tabular}{llcccc}
    \toprule
    \textbf{Sparse} & \textbf{Dense} & \textbf{MAE} & \textbf{$\text{MAE}_\lambda$} & \textbf{SAM} & \textbf{$\text{SAM}_\lambda$} \\
    \midrule
    True   & True        & 0.0448 & 0.0373 & 11.10 & 6.24 \\
    COLMAP & True        & 0.0414 & 0.0394 & 16.36 & 6.77 \\
    True   & Marigold-DC & 0.0805 & 0.0665 & 13.40 & 10.03 \\
    COLMAP & Marigold-DC & 0.0469 & 0.0415 & 11.45 & 7.53 \\
    \bottomrule
\end{tabular}
\end{table}

A potential avenue for future work is the joint refinement of distance maps within our recovery pipeline. By mutually regularizing appearance and geometry, such a formulation could significantly improve robustness to distance inaccuracies under challenging environmental conditions.

\end{document}